\documentclass[lettersize,journal]{IEEEtran}
\usepackage{amsmath,amsfonts}
\usepackage{array}
\usepackage{textcomp}
\usepackage{stfloats}
\usepackage{url}
\usepackage{verbatim}
\usepackage{graphicx}
\usepackage{cite}
\usepackage{xcolor}
\usepackage{enumitem}
\usepackage{makecell}
\usepackage{threeparttable}
\usepackage{booktabs}
\usepackage{caption}
\usepackage{subcaption}
\usepackage[noend]{algpseudocode}
\usepackage{algorithmicx,algorithm}
\usepackage{mdframed}
\usepackage{multirow}
\usepackage{array}
\usepackage{pifont}

\newmdenv[
  linecolor=black,
  linewidth=1pt,
  leftmargin=1mm,
  rightmargin=1mm,
]{myboxedtext}

\newcommand{\ouralg}[0]{PICE}

\newcommand{\ie}{\emph{i.e.}}
\newcommand{\eg}{\emph{e.g.}}
\newcommand{\etal}{\emph{et al.}}

\newcommand{\tann}[1]{\textcolor{black}{#1}}

\begin{document}


\title{\ouralg: A Semantic-Driven Progressive Inference System \\for LLM Serving in Cloud-Edge Networks}

\author{Huiyou Zhan, Xuan Zhang, Haisheng Tan,~\IEEEmembership{IEEE Senior Member}, Han Tian, Dongping Yong, \\ Junyang Zhang, and Xiang-Yang Li, ~\IEEEmembership{IEEE/ACM Fellow}

\thanks{Huiyou Zhan, Xuan Zhang, Haisheng Tan, Dongping Yong, Junyang Zhang and Xiang-Yang Li are with the LINKE Lab and the CAS Key Laboratory of Wireless-Optical Communications, University of Science and Technology of China (USTC), Hefei
230027, China (e-mail: zhanhuiyou@foxmail.com; xuanzhang@mail.ustc.edu.cn; hstan@ustc.edu.cn; yongdongping@mail.ustc.edu.cn; zhangjunyang@mail.ustc.edu.cn; xiangyangli@ustc.edu.cn).}
\thanks{Han Tian is with the University of Science and Technology of China (USTC), Hefei 230027, China (e-mail: henrytian@ustc.edu.cn).}}


\maketitle
\begin{abstract}

Large language models (LLMs), while driving a new wave of interactive AI applications across numerous domains, suffer from high inference costs and heavy cloud dependency. Motivated by the redundancy phenomenon in linguistics, we propose a progressive inference paradigm over cloud and edge, i.e., firstly generating the sketch of the answer by LLMs at cloud, and then conducting parallel extension to fill in details by small models (SLMs) at edge. Progressive inference offers potential benefits to improve throughput and reduce inference latency while facing key implementation challenges, including decreased response quality from SLMs, a tradeoff between the brevity and comprehensiveness of sketches, as well as increased latency caused by network transmission and edge inference. In this work, we propose and implement \ouralg, an LLM serving system with semantic-level cloud-edge collaboration, enhancing inference throughput and quality through dynamic inference task scheduling, ensemble learning, and parallel edge inference. Extensive testbed experiments illustrate that our approach achieves $1.5-2\times$ throughput enhancement and up to 43\% latency reduction, while also potentially enhancing the quality compared to SOTA systems.

\end{abstract}

\section{Introduction}
Large-scale language models (LLMs) such as GPT4, LLaMA, and Qwen \cite{achiam2023gpt, touvron2023llama, bai2023qwen} have exhibited unprecedented performance in natural language processing and chatbot systems. However, the exceptional capabilities of LLMs come at a significant inference cost, taking 22 seconds for Claude \cite{claude} (accessed through Slack API) and 43 seconds for Vicuna-33B \cite{chiang2023vicuna} (running locally on one NVIDIA A100 GPU) to generate a response \cite{ning2024skeleton}. 
To efficiently serve LLMs has emerged as a crucial demand for LLM-based AI applications to provide an engaging user experience, as their interactive nature mandates low response latency and high throughput during inference.

\begin{figure}[htbp]
    \centering
    \includegraphics[width=0.98\columnwidth]{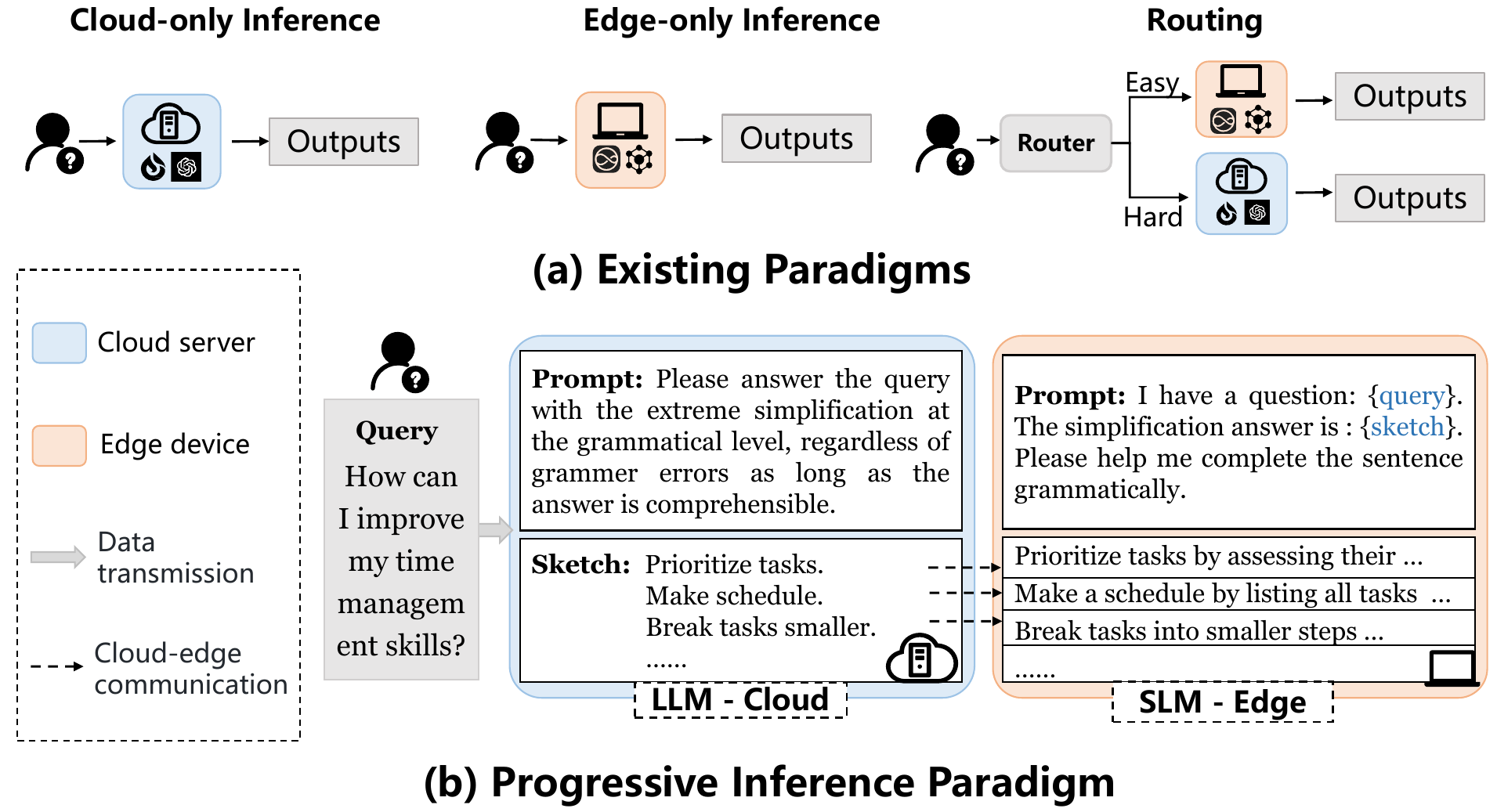}
    \caption{Existing inference paradigms vs. progressive inference.} 
    \label{fig:example}
\end{figure}

Moreover, due to the massive size of LLMs, current systems heavily rely on cloud computing  (Fig. \ref{fig:example}(a) - left), suffering from heavy request loads, high operational costs, and privacy concerns~\cite{zhang2024edgeshard}. To mitigate these issues, there has been a growing interest in designing smaller, cost-effective LLMs that can take advantage of the increasing availability of numerous and diverse edge computing devices. For instance, Meta provides multiple versions of Llama \cite{touvron2023llama} with only 7 billion parameters small enough to run on edge devices (Fig. \ref{fig:example}(a) - middle). Although this approach can enhance efficiency, it typically compromises response quality. To balance the need for fast and cost-effective inference, a hybrid inference strategy has been proposed, utilizing a router to dynamically allocate queries to either a small model on the edge or a large model on the cloud based on the predicted query difficulty \cite{ding2024hybrid, ong2024routellm, zhang2024treacle} (Fig. \ref{fig:example}(a) - right). However, this coarse-grained scheduling method is overly reliant on the performance of the router, highlighting the need for a more generalized approach that effectively harmonizes inference efficiency with response quality.


In contrast to existing efforts on optimizing inference at the model and system levels, we take a ``semantic-level” pathway to improve throughput by reducing the output of LLMs, \ie, \emph{the LLM creates the response sketch and the SLM fills in the details}. This perspective stems from how humans answer questions, \ie, we can convey the general meaning of a sentence with just a few words. 

Motivated by this idea, we propose a progressive inference paradigm over cloud and edge (Fig. \ref{fig:example}(b)). This architecture leverages the redundant computing power on the edge to offload workloads from the cloud while ensuring high response quality. Specifically, upon receiving a query, the LLM first evaluates the potential length of the response. If the response is concise and short, LLM will directly output a complete answer. In cases where a lengthy response is anticipated, LLM will produce semantically complete yet minimally grammatical sketches and then SLMs on the edge perform extension to improve inference efficiency. The semantic completeness of each short sentence eliminates sequential dependencies for their expansions, enabling SLMs to process each sentence in parallel and thereby breaking the sequential decoding of autoregressive models.

 The progressive inference paradigm is promising while introducing several key challenges in practical implementation: 1) The quality of responses from SLMs may decline; 
 2) The tradeoff between the brevity and the quality of LLM's sketches, \ie, while shorter sketches can improve throughput, the semantics expressed may be incomplete; and 3) Additional latency caused by network transmission and edge inference. 

In this work, we present \ouralg, an efficient progressive inference system of LLMs, which synergistically makes use of the resources at the cloud and edge, effectively reducing the workload of LLMs on the cloud, and achieving both high throughput and superior response quality. The effectiveness of \ouralg~stems from four critical designs: a dynamic scheduler taking advantage of progressive inference to optimize throughput, an ensemble learning mechanism aggregating outputs from multiple small models to improve quality, a parallel execution optimization strategy for SLMs to speed up edge inference and a specialized fine-tuning component that enables LLMs to generate concise yet semantically complete sketches. Our main contributions are highlighted as follows.

\begin{itemize}[leftmargin=*]
    \item We propose and demonstrate the feasibility of synergistic progressive inference at the semantic-level parallel data processing across the cloud and edge. To the best of our knowledge, this is the first attempt to involve semantic optimization with cloud-edge collaboration to enable efficient and reliable LLM serving (Sec. \ref{sec:motivation}).
    \item We design \ouralg, a novel \underline{P}rogressive \underline{I}nference engine for LLM serving over \underline{C}loud and \underline{E}dge, achieving high throughput and low latency by employing dynamic scheduling strategies, ensemble learning mechanism and parallel edge inference (Sec. \ref{sec:overview} and Sec. \ref{sec:design}). 

    \item By implementing  \ouralg~at a testbed with a set of Jetson AGX Orins and a cloud server, we conduct extensive experiments across various LLM benchmarks and popular language models, demonstrating that \ouralg~achieves a throughput enhancement of 1.5$\times$ to 2$\times$ and a latency reduction by up to 43\%, while potentially enhancing the answer quality on several question categories compared to baselines (Sec. \ref{sec:evaluation}). 
\end{itemize}

\section{Background and Motivation}\label{sec:motivation}

    

\subsection{LLM Services on the Cloud and Edge}
As shown in Table \ref{tab:ModelPerformance}, the massive size of LLMs, while offering high response quality, entails significant computational resources that can only be met by cloud servers \cite{yu2022orca}. As more LLM-based services are introduced, the exclusive reliance on the cloud is increasingly unable to meet the requirements for inference efficiency such as the throughput and latency. 

\begin{table}[tbhp]
\centering
\captionsetup[table]{singlelinecheck=off}
\begin{threeparttable}
\begin{tabular}{cccc}
 \toprule
 Model & Speed(tokens/s)& GPU Memory(GB) & MMLU\\
  \midrule
    Qwen2.5-72B-Instruct & 18.19 & 134.74 & 86.1\\
    Llama3-70B-Instruct & 18.82 & 130.64 & 79.5\\
    Qwen2.5-32B-Instruct  & 22.13 & 60.11 & 83.3\\
    Llama3-8B-Instruct & 76.5 & 15.83 & 66.6\\
    Qwen2.5-7B-Instruct & 84.28 &  14.92 & 74.2\\
    Qwen2.5-1.5B-Instruct & 183.33 & 3.44 & 60.9\\
 \bottomrule
\end{tabular}
\end{threeparttable}
\caption{Performance Comparison. The speed of the models is tested on two NVIDIA A100 GPUs with vLLM framework. The MMLU score is a metric for assessing the performance of language models in multi-task language understanding \cite{hendrycks2020measuring}.} 
\label{tab:ModelPerformance}
\end{table}

Edge computing is a promising solution to address the aforementioned challenges by providing service on edge devices closer to the users~\cite{hua2023edge}. 
Constrained by computational power and memory, the size of the language models deployed on the edge is typically much smaller than those on the cloud, leading to a decrease in their response quality. For instance, 8B parameters represent the maximum size for Llama to achieve real-time inference on smartphones (\eg, Xiaomi 10) \cite{xu2023llmcad}, yet its arithmetic reasoning capacity is significantly lower than that of the Llama3-70B model (Table \ref{tab:ModelPerformance}).

Faced with the tradeoff between inference efficiency and response quality, recent works have introduced a new inference paradigm called routing, which uses two models of different sizes instead of a single model for inference \cite{ding2024hybrid, srivatsa2024harnessing}. The smaller model generally has lower inference costs but also lower accuracy than the larger one. The key idea is to identify and route easy queries to the small model so that inference cost can be reduced while maintaining response quality. These approaches typically require training routers from scratch for each scenario, which may lack generalizability. There are also model-level and system-level inference optimization techniques \cite{lin2024infinite}, such as model cascading \cite{wang2023tabi}, speculative decoding \cite{leviathan2023fast}, PagedAttention\cite{kwon2023efficient}, and continuous batching\cite{yu2022orca, yang2024queueing}, which are mostly applicable to cloud servers and do not fully leverage the computing power of edge devices. The continuous growth of edge intelligent devices motivates us to design a more general cloud-edge LLM serving paradigm.

\subsection{Our idea: Cloud-edge Collaborative Progressive Inference}

The effectiveness of Transformer-based LLMs is hindered by the quadratic increase in computational cost relative to sequence length, especially when dealing with long sequences. 
Due to the autoregressive nature of LLMs, generating one token would require reading the entire KV cache. For the Llama3-8B model with 32k context length, the KV cache can occupy 16GB of space, requiring at least 11 ms for reading, which contributes to more than 50\% of the inference latency and limits the overall throughput \cite{tang2024quest}. This inspires us to investigate the potential for enhancing inference efficiency through reducing the LLM's response lengths. The feasibility of our idea is supported by the following two observations.

\begin{figure}[htbp]
\centering
    \includegraphics[width=0.9\columnwidth]{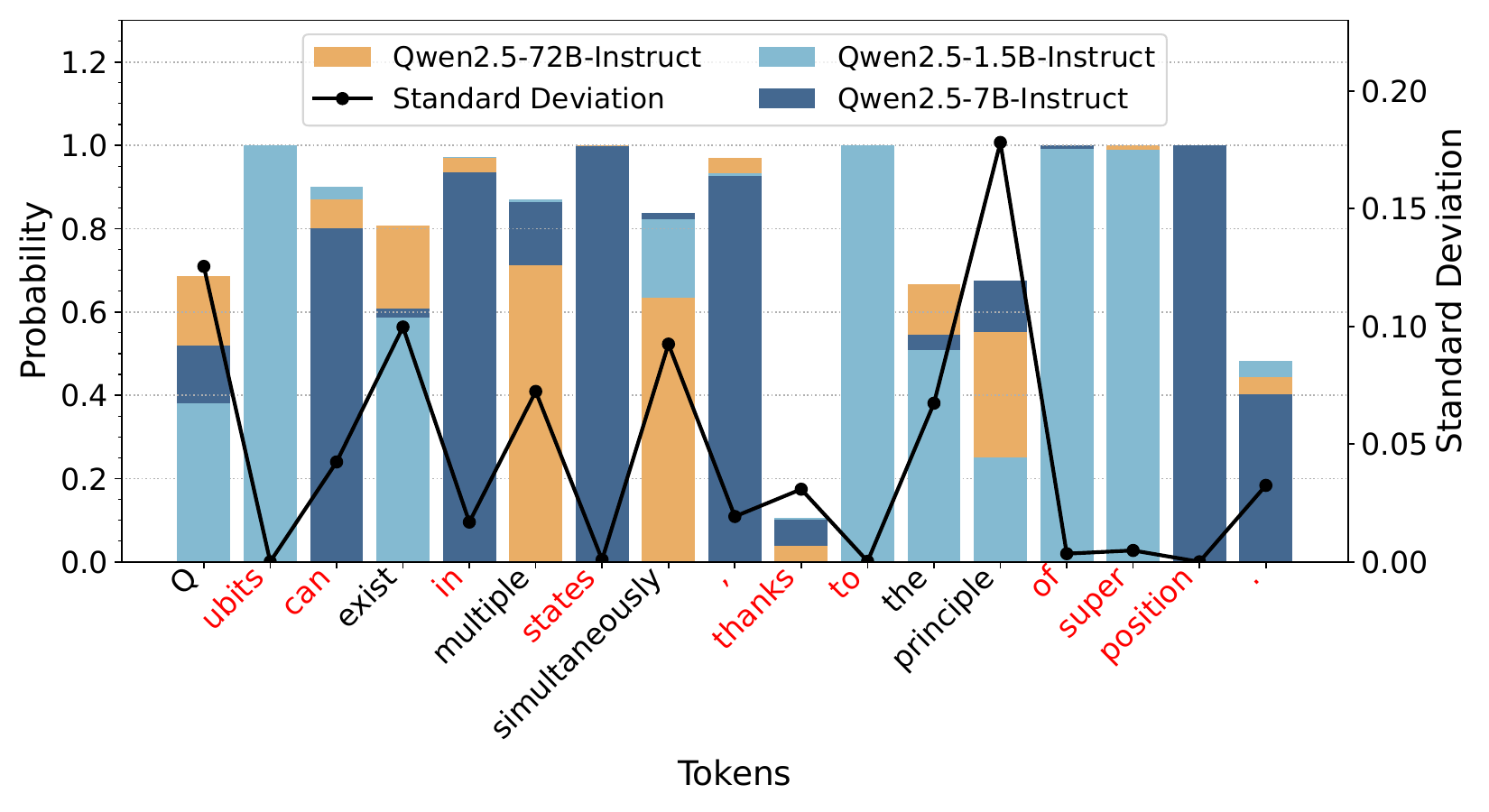} 
        \caption{The conditional probability and variances of the Qwen2.5-72B, Qwen2.5-7B, and Qwen2.5-1.5B models across different tokens. Lower variances indicate closer agreement among the models' output distributions.}
        \vspace{-0mm}\label{mov:ob1}
\end{figure}

\noindent\textbf{Observation 1: Not all tokens play the same role in expressing semantics.} As the redundancy phenomenon in linguistics, the semantic essence of a sentence is often conveyed by just a few words, with others only to refine the grammatical structure. This principle is also evident in the inference of LLMs. For instance, previous work noted that in sentiment analysis tasks, tokens that express strong emotions carry the highest weights in the vector of importance, whereas adpositions such as "and" and "to" receive the least attention \cite{wang2023tabi}. Typically, the performance differences among various models are manifested in these critical tokens. As illustrated in Fig. \ref{mov:ob1}, Qwen2.5-72B  exhibits the greater variance in probability distributions for tokens such as ``Q", ``exist", ``simultaneously", and ``principle", compared to its 7B and 1.5B counterparts.

\noindent\textbf{Observation 2: Once key tokens are given, both LLM and SLM exhibit similar conditional probabilities for other tokens.} We analyze the conditional probability distributions of the models when generating the same answer. As illustrated in the Fig. \ref{mov:ob1}, after providing several key tokens (\eg, Q, exist, simultaneously), the variance in the models' conditional probability distributions for the remaining tokens is relatively low, especially for prepositions like ``to" and ``of". This suggests that if the LLM can provide the SLM with key tokens, the SLM's response quality can be similar to the LLM's.

\begin{figure}[htbp]
\centering
    \includegraphics[width=0.85\columnwidth]{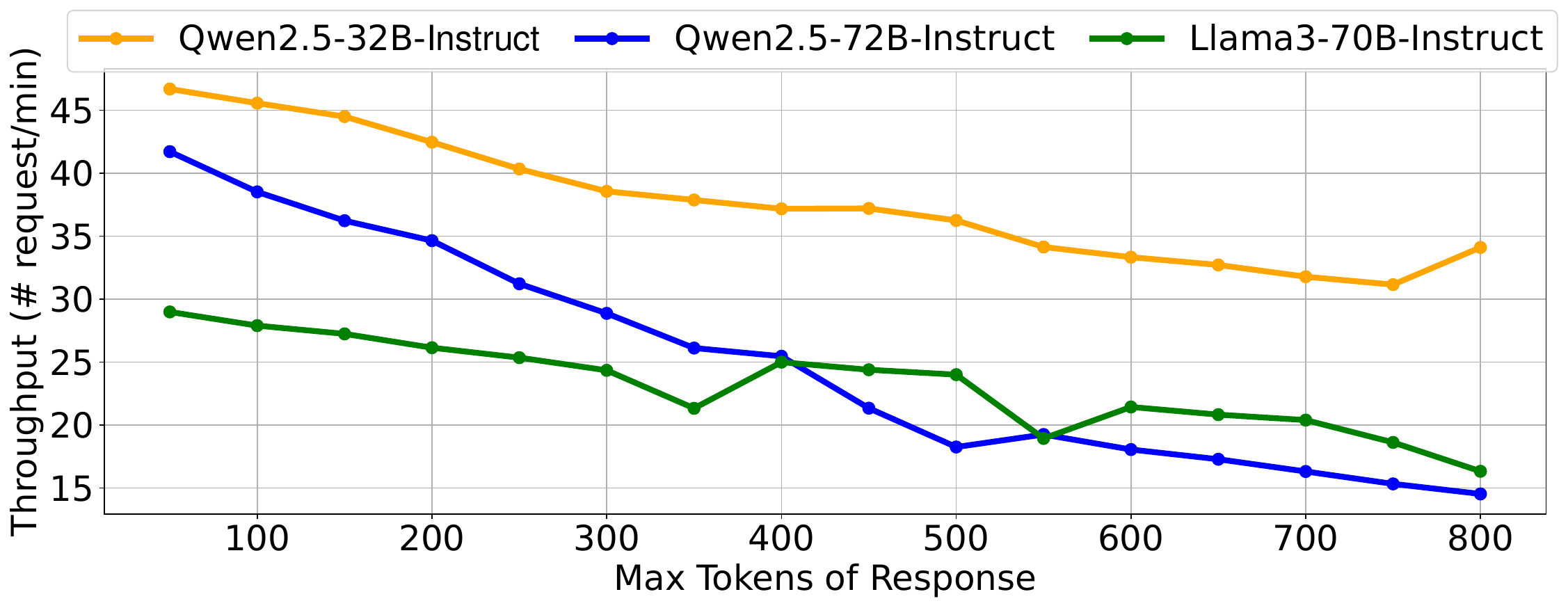} 
        \caption{The horizontal axis represents the max tokens of the LLM's response, while the vertical axis denotes the throughput of the serving system. The throughput is measured in the number of queries processed per minute (\#queries/min).}
        \label{mov:performance}
\end{figure}

Drawing from the above observations, we adopt progressive inference to increase throughput by shortening the responses of LLMs. Specifically, leveraging the response length awareness \cite{zheng2024response}, the LLM addresses long-answer questions through extreme grammatical simplification, as long as the answer is comprehensible regardless of grammatical errors. The simplified answer is subsequently provided to the SLM at edge to be further expanded as a full response. Since each sentence of the sketches is semantically complete, the expansion of different sentences becomes independent. This independence enables SLMs to expand sentences in parallel, breaking autoregressive decoding and thus greatly speeding up the inference process. \tann{According to our preliminary experiments  (Fig. \ref{mov:performance}),} by employing progressive inference, the generation length of the LLM can be reduced from an average of 500 tokens to 200, thereby increasing the throughput by 1.5-2$\times$.

\subsection{Challenges of Progressive Inference}\label{sec:2.3}


\noindent\textbf{Challenge 1: Ensuring high response quality.} There is a risk of declined response quality due to the limited inference capability of SLMs, particularly when the sketches provided by the LLM are excessively concise. Ensuring that the SLM can effectively expand and enrich the LLM’s skeletal responses without compromising accuracy or coherence poses a significant challenge.

\noindent\textbf{Challenge 2: Balancing response length and quality.} A central challenge involves optimizing the trade-off between the length and quality of sketches. As shown in Fig. \ref{mov:performance}, shorter sketches by the LLM lead to higher throughput, resulting in fewer tokens for the SLM to utilize as references, potentially reducing response quality. This issue underscores the careful consideration of how much information should be passed from the cloud to the edge.

\noindent\textbf{Challenge 3: Handling latency due to network transmission and edge inference.} The integration of network transmission and edge inference adds another layer of complexity. Even though progressive inference only requires transmitting minimal text, network congestion can still adversely affect performance. Additionally, the limited computational capability of edge devices calls for efficient dynamic scheduling of tasks. 

Addressing the above challenges requires efforts to enhance the collaborative dynamics between models and optimize the data flow across the cloud and edge, which is crucial for harnessing the full potential of the progressive inference paradigm for a more efficient and responsive inference system. 

\section{\ouralg ~Overview}\label{sec:overview}

\begin{figure}[tbhp]
    \centering
    \includegraphics[width=0.99\columnwidth]{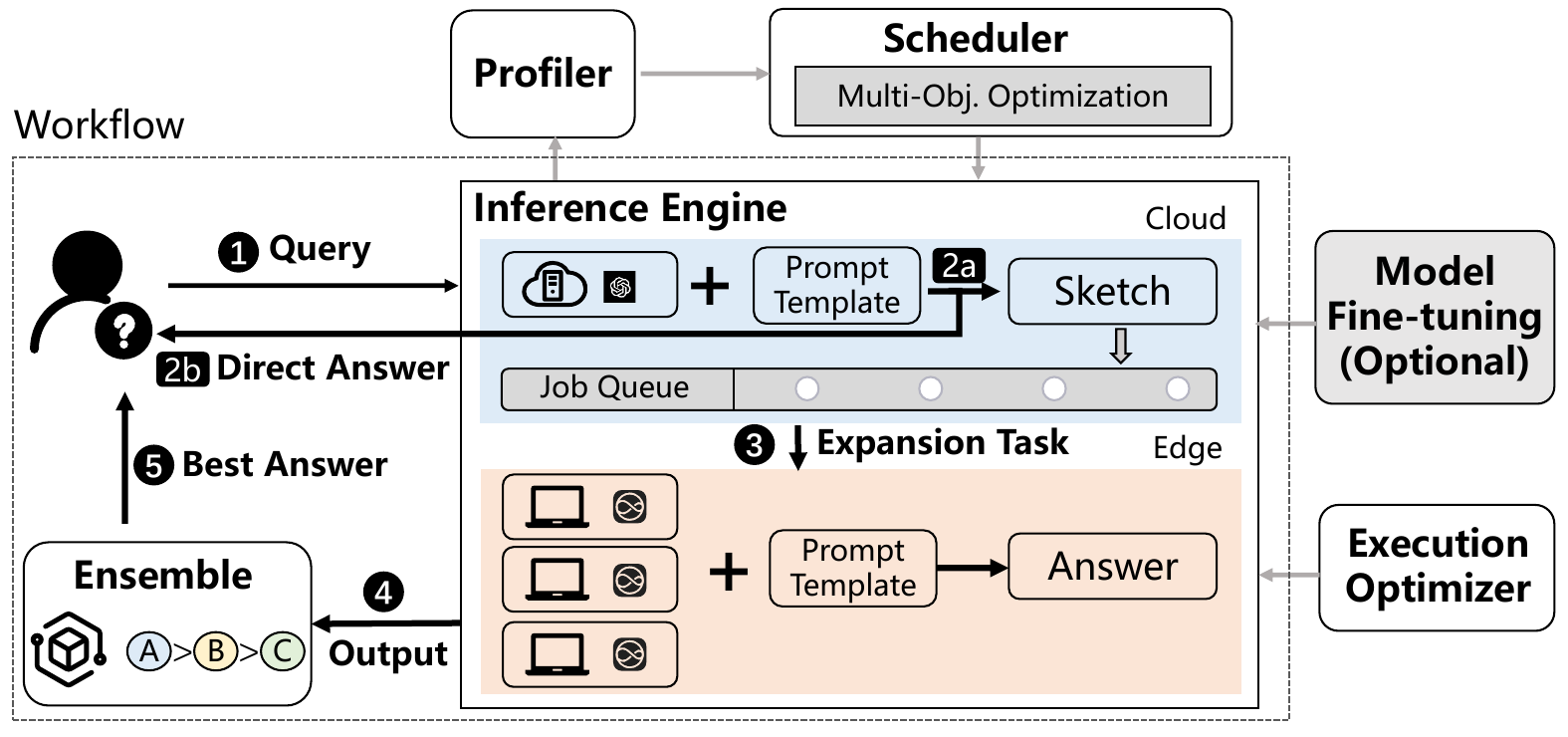}
    \caption{
        The overview and workflow of \ouralg.
    } 
    \label{fig:algoverview}
\end{figure}

Motivated by the huge overheads of serving LLMs and the new optimization opportunities brought by progressive inference, \ouralg~proposes a novel cloud-edge collaborative inference engine to achieve fine-grained throughput improvement efficiently for generic models driven by real-time per-query feedback. 
Fig.~\ref{fig:algoverview} shows a high-level view of \ouralg~and its workflow. 
The cloud-based LLM inference engine assesses the response length of incoming user requests (\ding{182}), providing a full answer directly (\ding{183}a), or generating a sketch for complex responses and queuing an expansion task (\ding{183}b). For the latter, the scheduler assigns this task to appropriate edge models (\ding{184}), and then the system selects the output answer (\ding{185}) with the highest confidence to the user (\ding{186}). We next introduce the components of \ouralg~in details. 

\noindent\textbf{Scheduler.} The control part of \ouralg~is a scheduler that capitalizes on the advantages of progressive inference to boost system throughput.
It dynamically adjusts the sketch length generated by the cloud-based LLM based on complex runtime conditions, intelligently allocates tasks between the cloud and edge to optimize resource utilization, and selects optimal model candidates at the edge for task execution.

\noindent\textbf{Ensemble Learning Mechanism.} To offset potential declines in response quality, our system implements an ensemble learning strategy, where multiple smaller models collaborate to generate and select the highest quality response. This method leverages the collective intelligence of diverse models to enhance decision accuracy.

\noindent\textbf{Execution Optimizer.} Designed to accelerate inference on edge devices, the execution optimizer is crucial for refining the expansion strategy of sketches. It ensures uniform sequence lengths within batches to maximize processing parallelism, significantly improving overall system efficiency.

Besides the above three key components, there are other components incorporated within our LLM service system.
\begin{itemize}[leftmargin=*]
    \item \noindent\textbf{Profiler.} It operates in two phases: offline and runtime. In the offline phase, it conducts device-specific latency estimation. During runtime, it continuously monitors device and server loads, as well as network conditions, providing real-time data to the scheduler for informed decision-making.
    \item \noindent\textbf{Inference Engine.} The engine is responsible for executing AI model inferences, divided into cloud and edge parts, and incorporates prompt templates for progressive inference. The cloud component runs LLMs, while the edge component utilizes SLMs across multiple devices for faster computation and improved response time. A job queue buffers tasks from the LLM, which are then optimally distributed by the scheduler to the edge models for processing.
    \item \noindent\textbf{Model Fine-tuning (Optional).} To improve the quality of sketches generated by the LLM, we include an optional fine-tuning component. This component uses reinforcement learning from AI feedback to optimize the LLM for creating concise sketches that capture complete semantics with minimal wording. The fine-tuning is performed offline, and the optimized model is then deployed to the cloud for use in cloud-edge progressive inference.
\end{itemize}

In summary, our system \ouralg~adopts a semantic-level inference acceleration technique that works with existing models and hardware without any changes, which is also compatible with optimization techniques such as speculative decoding \cite{leviathan2023fast}, PagedAttention~\cite{kwon2023efficient}, and continuous batching\cite{yu2022orca}. 

\section{Key Design} \label{sec:design}

The core of \ouralg~is a novel progressive inference engine, employing efficient and powerful models to serve queries. In this section, we introduce key designs that fully harness the potential of progressive inference.

\subsection{Dynamic Scheduler}

\subsubsection{Optimization Objective}
We first present the optimization objectives of the dynamic scheduler and formalize the scheduling problem. Let us consider a dataset \( D = \{(x_i, y_i)\}_{i=1}^{N} \) made up of \( N \) pairs of inputs \( x_i \) and their corresponding outputs \( y_i \). To address the modeling of the transformation from \( x \) to \( y \), we employ a two-stage modeling strategy. Firstly, the input \( x \) is transmuted into a more meaningful sketch \( r \) by a LLM \( h \), symbolized as \( h: x \rightarrow r \). Following this, the subsequent phase consists of employing an SLM \( s \), which takes the extracted representation \( r \) and forecasts the output \( y \), expressed as \( s: r \rightarrow y \). Through this approach, we essentially break down the complete \( x \rightarrow y \) modeling operation as $p(\hat{y}|x) = p_h(r|x) \cdot p_s(y|r)$. Here, \( p(\hat{y}|x) \) is the probability model of \ouralg, serving as an approximation of \( p(y|x) \) with error $\delta(\hat{y}, y)$ ($\delta$ is the error function).

Therefore, one of our goals is to decrease the error, represented by \(E[\delta(\hat{y}, y)]\), through the dual optimization of the SLM \(s\) and the sketch \(r\): 
\begin{equation}
\sigma^{*} = \mathop{\arg\min}\limits_{s \in S, r \in R} E[\delta(\hat{y}, y)] = \mathop{\arg\min}\limits_{s \in S, r \in R} E[\delta(s(h(x)), y)]  \label{eq:2}
\end{equation}
where \(\sigma^{*}\) denotes the optimal SLM and sketch over all models in set \(S\) and all possible sketches $R$.

Another objective of the system is to maximize throughput. We define the average time $f(l)$ for the cloud-based LLM to process a response of length $l$. Let $c$, the cost coefficient, be the ratio between the time for a single execution of the LLM in the cloud and the SLM at the edge. The value of $c$ is not only dependent on the model but also on hardware and software details. Consequently, the average processing time for the SLM to generate a response of length $l$ is $c \cdot f(l)$. The throughput can be formulated as $\max( f(|r|), \Delta(r), \frac{c}{p} \cdot f(l) )$, where \( \Delta(r) \) is the network latency between edge and cloud, and $p$ denotes the degree of parallelism for edge inference. Additional constraints include maintaining the processing latency for each request below $f(l)$, the latency for cloud inference.

To enable the support of realistic multi-criteria service level agreements (SLAs), we model our scheduling problem as a multi-objective optimization problem. The current set of metrics is defined as 
\begin{equation}
M = \{\text{error}, \text{throughput}, \text{latency}, \text{server cost}, \text{edge cost}\} \notag
\end{equation}
We define server cost and edge cost based on the count of generated tokens, respectively. Similar to SPINN \cite{laskaridis2020spinn}, we categorize the aforementioned objectives into hard constraints, such as inference latency, and soft constraints, such as maximizing throughput and minimizing errors. To enable the user to specify the importance of each metric, we employ a multi-objective lexicographic formulation \cite{marler2004survey}:
\begin{align*}
\min_{\sigma} M_i(\sigma), \quad s.t. \quad M_j(\sigma) \leq M_j(\sigma^*_{j}) \\
j=1,2,...,i-1, \ i > 1, \ i=1,2,...,|M|
\end{align*}
where \( M_i \in \{M_1,M_2,...,M_|M|\} \) is the \( i \)-th metric in the ordered tuple of soft targets, \( i \) is the position of the metric in the importance sequence, and \( M_j(\sigma^*_{j}) \) represents the optimum of the \( j \)-th metric found in the \( j \)-th iteration. Under this formulation, we rank the metrics according to their importance as required by the target use case.

\subsubsection{Cloud-side scheduling}

We address this optimization problem through multiple steps. When conducting LLM inference in the cloud, we first ensure that the hard constraint of inference latency is met, while enhancing throughput and response quality. Cloud-side scheduling primarily achieves this by dynamically adjusting the length of the sketch.

Specifically, the total end-to-end latency of a query comprises the latency components from LLM inference, waiting time, network transmission, and SLM inference. We define the LLM inference latency as a function of the output length, \( f \), such that the inference latency for a task \( r_i \) on the SLM is given by \( \frac{c \cdot f(l_i)}{p} \), where \( l_i \) denotes the expected response length for the query. The waiting time represents the total completion time for all tasks in the job queue \( Q \) on the edge, calculated as \( \frac{\sum_{r_j \in Q} c \cdot f(l_j)}{Np} \), with \( N \) representing the number of edge devices. Thus, the constraint for the end-to-end latency can be formulated as follows:

\begin{equation}
f(|r_i|) + \Delta(r_i) + c\cdot f(l_i) + \frac{\sum_{r_j \in Q} c \cdot f(l_j)}{pN} \le f(l_i) \label{eq:4}
\end{equation}

Our scheduling algorithm is inspired by the observation that, similar to humans' ability to estimate the length of an answer based on their understanding of a query, LLMs can also perceive the length of responses in advance, as demonstrated by previous work \cite{zheng2024response}. In our workflow, the LLM outputs a sketch along with the expected answer length $l_i$. In our workflow, the LLM outputs a sketch and the expected answer length \( l_i \). Leveraging the measurements of \( f \) and \( c \) that we established during the offline stage with our profiler, we can estimate the edge inference latency and the waiting time.

We define multiple sketch length levels, from 0 to \( l_i \). Upon receiving a query, we first evaluate the network's conditions and estimate edge inference latency conservatively, setting \( p=1 \) by default. If no level above 0 meets the criteria set by inequality (\ref{eq:4}), we forgo progressive inference and directly request a complete response from the LLM. If the conditions permit, we assess the SLM's ability to select a suitable sketch length, with more capable SLMs potentially opting for shorter lengths to increase throughput. Note that since the sketch length is specified via LLM prompts, \( r_i \) may differ by up to 10 tokens from the expected value.

\subsubsection{Job dispatching}

In the process for handling queries that require progressive inference, it is crucial to allocate expansion tasks among edge devices in a manner that ensures load balancing. To manage the dynamic nature of these workloads, we implement a multi-list scheduling algorithm that adaptively allocates inference tasks according to the varying demands.

In practical scenarios, response lengths to different queries can vary significantly. This variability, when batching SLM inferences, introduces inefficiencies due to the inclusion of sequences with disparate lengths. Specifically, shorter sequences must wait for longer ones to complete, leading to computational delays. The quadratic time complexity of the inference further exacerbates these inefficiencies, placing a substantial load on the process.

Our job dispatching algorithm, delineated in Algorithm \ref{alg:job_dispatching}, is primarily a multi-list scheduling algorithm that categorizes tasks based on the expected answer length \( l_i \). As queries are received into the job queue \( Q \), they are assigned to specific lists based on their length (Lines 2-6). Once edge devices are idle, they proactively retrieve a batch of tasks from the longest list to perform inference (Lines 7-11).

\begin{algorithm}[htbp]
\caption{Job Dispatching}
\label{alg:job_dispatching}
\begin{algorithmic}[1]
    \State Initialize lists $q_1, q_2, \ldots, q_n$ for different lengths.
    \State $Q \gets$ List of jobs ($r_i, l_i$) waiting for processing
    \While{$Q$ is not empty}
        \State Remove task $r_i, l_i$ from $Q$
        \State Determine list index $j$ based on $l_i$
        \State Add $r_i$ to $q_j$
    \EndWhile
    \For{each edge device $D$}
        \While{$D$ is idle}
            \State $max \gets$ List with the most jobs
            \State $batch \gets$ SelectBatchFrom($q_{max}$)
            \State $D$.ProcessBatch($batch$)
        \EndWhile
    \EndFor
\end{algorithmic}
\end{algorithm}

\subsubsection{Edge-side scheduling}

To cope with the tradeoff between efficiency and response quality, the scheduler must select appropriate SLMs on edge devices for inference. The model selection process includes offline profiling and online selection. Profiling generates a range of differently optimized candidates. When a task arrives, \ouralg~selects a candidate that can meet the targets with the lowest latency.

\noindent\textbf{Offline profiling.} The performance of a candidate is jointly decided by the base models and hardware. We initially utilize the profiler to assess the inference latency of candidate SLMs and LLMs when generating sequences of a specified length on edge devices and cloud servers, respectively, and calculated their cost coefficient \( c \). Subsequently, we test the inference latency of the LLM on the cloud server when generating sequences of varying lengths. Both results serve as a basis for estimating the execution time of tasks in the job queue.

\noindent\textbf{Online candidate selection.} As shown in Algorithm \ref{alg:model_selection}, upon an edge device retrieving a task \( r_i \) from the job queue, it first estimates the remaining processing time based on the LLM's anticipated response length \( l_i \) and offline profiling data (Line 2). If the total inference time for the task exceeds the threshold \( f(l_i) \), thereby violating the hard constraint of the optimizer, a smaller SLM must be selected (Lines 3$\sim$4). Conversely, if the total inference time is less than the threshold, a higher quality SLM may be chosen for inference or the current SLM can be retained. To avoid the overhead associated with frequent model switching, the latter process is only conducted when the number of pending tasks in the job queue is less than its maximum (Lines 6$\sim$12).

\begin{algorithm}[hbpt]
\caption{Model Selection}
\label{alg:model_selection}
\begin{algorithmic}[1]
    \State \textbf{Input:} $r_i, l_i, f(l_i)$, $f(|r_i|)$
        \State Estimate remaining processing time $\tau_i$ for current SLM.
        \If{$\tau_i > f(l_i) - f(|r_i|)$}
            \State Switch to a smaller SLM for $r_i$
        \Else
            \If{$|JobQueue| < \text{maximum}$}
                \For{each SLM $m_j$ larger than the current SLM}
                    \State Calculate $\tau'_i$ for $r_i$
                    \If{$\tau'_i < f(l_i) - f(|r_i|)$}
                        \State Upgrade to $m_j$ for $r_i$ and break the loop
                    \EndIf
                \EndFor
            \Else
                \State Continue with current SLM for $r_i$
            \EndIf
        \EndIf
\end{algorithmic}
\end{algorithm}

\subsection{Execution Optimizer}

Despite the reduced model size on edge devices, the inference latency, limited by their computational and memory capacities, can still present a significant bottleneck for the overall system. Hence, it is crucial to expedite the inference process on edge devices.

We optimize edge inference through semantic-level parallelism strategies. Specifically, the sketch received by the edge inference engine consists of multiple short sentences. As each short sentence is semantically complete, their expansions are independent of each other and can therefore be executed in parallel. The prompt template is as follows.
\begin{myboxedtext}
$[\text{System}]$: I have a question about \{\textbf{query}\}. The simplification answer is as follows: \{\textbf{sketch}\}. 

\noindent Now, please help me complete and only complete the writing of a short sentence \{\textbf{sentence}\}. Do not continue with other sentences!

\noindent $[\text{Assistant}]$: ......
\end{myboxedtext}
However, higher parallelism is not always preferable. This is due to two factors. (1) Sentence length variability. The differing lengths of sentences can lead to unnecessary waiting when simply batched for processing. (2) Prompt overhead. The need to provide additional sketch information can result in longer prompts for the SLM to handle. 
Excessive parallelism may result in redundant sketch information in the KV cache, reducing system efficiency.

To tackle the challenges, we adopt a binary-tree-based merging strategy for short sentences. Considering a sketch \( r_i \) with \( k \) short sentences $( r_{i,1}, r_{i,2}, \ldots, r_{i,k} )$, we assume that the number of words in a short sentence is positively correlated with the number of words after its expansion. Accordingly, we sort the \( k \) short sentences by their word count and combine them into \( \lceil \frac{k}{2} \rceil \) groups as follows: \( (r_{i,1}, r_{i,k}), (r_{i,2}, r_{i,(k-1)}), \ldots (r_{i,\lfloor \frac{k+1}{2} \rfloor}, r_{i,\lceil \frac{k+1}{2} \rceil})\). If the current degree of parallelism can still satisfy the hard constraint of inference latency, we recursively apply the aforementioned merging process to increase the throughput of edge devices.

\subsection{Ensemble Learning}

The open-source SLMs exhibit diverse strengths and weaknesses due to variations in data, architectures, and hyperparameters, making them complementary to each other. This diversity motivates us to dynamically ensemble these SLMs to generate consistently better responses for each input. A typical ensembling approach involves selecting the optimal candidate from outputs generated by different models, utilizing evaluation metrics such as MLM-Scoring \cite{salazar-etal-2020-masked} and BERTScore \cite{zhangbertscore} to train a reward model \cite{jiang2023llm}. However, these methods are inapplicable to our system due to the additional overhead caused by training and deploying the reward model. To avoid the overhead caused by the ensemble, a straightforward way to get the confidence is to use perplexity \cite{jelinek1977perplexity}, a metric for assessing a model's ability to understand and generate human-like language. Perplexity is calculated as the exponential of the average negative log-likelihood of a word sequence.

However, we have observed in practice that perplexity is overly dependent on the model itself. For instance, the Llama3-8B model consistently exhibits higher perplexity across all queries compared to the Qwen2.5-7B, regardless of the actual quality of the responses. To holistically evaluate both the model and the quality of responses, we use a weighted average of text scores and perplexity as the confidence. The text score comprises two components: one is the length of the answer, where more detailed expansions by the SLM receive higher scores; the other is the similarity between the answer and the sketch, with closer matches scoring higher, represented by the rouge-l score \cite{lin2004rouge}. Thus, the confidence of a sequence can be calculated as follows.
\begin{equation}
\begin{split}
con(\hat{y}) = & \ \alpha_1\cdot 2^{\frac{1}{N} \sum_{i=1}^{N} \log_2 p(w_i)} + \alpha_2 \cdot \text{Norm}(|\hat{y}|) \\
& + (1-\alpha_1-\alpha_2)\cdot \text{Rouge-l}(r, \hat{y})
\end{split}
\label{eq:perplexity}
\end{equation}
where $N$ is total number of tokens, $w_i$ is the $i$-th token in the answer and  $p(w_i)$ is probability assigned by the model to the $i$-th token. The parameters \(\alpha_1\) and \(\alpha_2\) are weights, while \(\text{Norm}(|\hat{y}|)\) represents the normalization of the response length. To mitigate the impact of sampling strategies, a single SLM may produce more than one sequence.



\subsection{Model Fine-tuning}

\begin{figure}[tbhp]
    \centering
    \includegraphics[width=0.99\columnwidth]{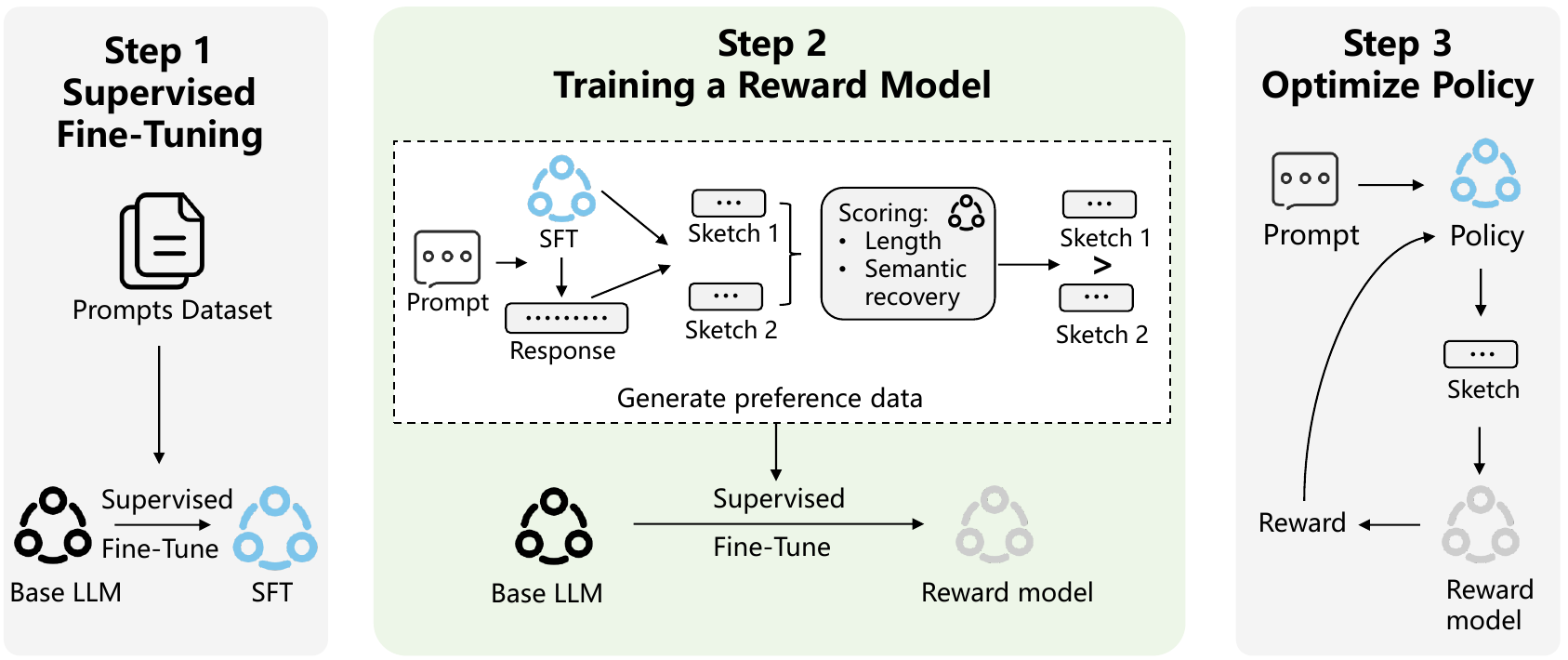}
    \caption{
        The design of \ouralg's fine-tuning component.
    } 
    \label{fig:rlhf}
\end{figure}

The model fine-tuning component in \ouralg~is an optional module designed to enhance the LLM's ability to generate concise sketches that convey complete semantics with minimal wording. Unlike typical reinforcement learning from human feedback (RLHF), this component introduces a specialized preference labeling algorithm tailored to refine the model’s ability to generate high-quality sketches. 
As shown in Fig.~\ref{fig:rlhf}, the fine-tuning process consists of three steps: supervised fine-tuning, reward model training, and reinforcement learning-based fine-tuning.

In the first step, the pre-trained LLM is fine-tuned on a labeled dataset to generate concise sketches (e.g., given an input document, generating a summary that captures the main idea of the document),  using token-level supervision to produce a supervised fine-tuned (SFT) model $\pi$.

In the second step, our goal is to generate a reward model that takes in a sequence of sketches, and returns a scalar reward which should numerically represent our preference for a good sketch. The training dataset of prompt-generation pairs for the reward model is generated by the SFT model and a preference labeling algorithm. Given an input \( x \), \ouralg~first obtains a full answer \( y \) from the SFT model \( \pi \). Then, the query \( x \) and answer \( y \) are provided to the model \( \pi \), which generates a pair of sketches \( (r1, r2) \) for \( y \). The input and sketches are sent to the preference labeling algorithm to rate which sketch is better according to the following criteria. 
\begin{itemize}
    \item The shorter the length of the sketch ($l_r$), the higher the score it receives. 
    \item The base LLM expands the sketch \( r \) into a full answer \( \hat{y} \), which is compared to the answer \( y \) from the SFT model. Higher similarity indicates a higher score for the sketch.
\end{itemize}
The score of a sketch can be calculated as follows: $\beta_1\cdot \frac{1}{l_r} + \beta_2\cdot\text{Rouge-l}(\hat{y}, y)$. The parameters $\beta_1$ and $\beta_2$ are weights. These elements form a triplet dataset \( D = \{(x, r_w, r_l)\} \), where \( r_w \) and \( r_l \) are the preferred (i.e., higher-scoring) and non-preferred sketches, respectively. The reward model (RM) \( R_\phi \) is trained by minimizing the following loss: 
\begin{align*}
\mathcal{L}_R(\phi)=\underset{\left(x, r_w, r_l\right) \sim \mathcal{D}}{-\mathbb{E}}\left[\log \sigma\left(R_\phi\left(x, r_w\right)-R_\phi\left(x, r_l\right)\right)\right],
\end{align*}
where $\sigma$ is the sigmoid function.

In the third step, a policy \( \pi_\theta^{RL} \) is initialized with the weights of the SFT model and then refined using reinforcement learning to maximize the reward from the RM, which serves as a proxy for sketch preferences. To prevent excessive divergence from the original SFT policy \( \pi \), a Kullback-Leibler (KL) divergence \cite{kullback1951information} term \( D_{KL} \) is incorporated into the objective. This term, controlled by the hyperparameter \( \gamma \) \cite{geist2019theory}, ensures that \( \pi_\theta^{RL} \) does not produce outputs that, although highly rewarded by the RM, have low quality or unnatural language \cite{amodei2016concrete}.
The optimization objective is defined as follows:  
\begin{equation}
\begin{split}
J(\theta) = & \mathbb{E}_{r \sim \pi_\theta(\cdot | x)} [ (1 - \gamma) R_\phi(r | x) \\
& - \gamma D_{KL}(\pi_\theta^{RL}(r | x) \| \pi(r | x)) ].
\end{split}
\notag
\end{equation}
\section{Evaluation} \label{sec:evaluation}
We evaluate \ouralg~through extensive testbed-based experiments with various LLM benchmarks and popular language models. We first assess its overall performance in enhancing inference efficiency and ensuring response quality (Sec. \ref{sec:methodology}). Next, we examine its three core components within the cloud-edge progressive inference framework ((Sec. \ref{sec:performance_breakdown})). Additionally, we analyze the impact of hyperparameters on system performance (Sec. \ref{sec:sensitivity_analysis}). The main takeaways are:
\begin{itemize}
    \item \ouralg~achieves $1.5\times\sim 2\times$ throughput improvement and up to 43\% latency reduction over baseline methods.
    \item \ouralg~effectively maintains response quality, and even enhances it on several question categories.
    \item  The strategic optimizations within the LLM serving system enable \ouralg~to effectively balance inference efficiency and quality in various scenarios.
\end{itemize}

\subsection{Methodology}\label{sec:methodology}

\noindent\textbf{Testbed setup.} Our experimental infrastructure comprises eight devices, including four Jetson AGX Orin units and a cloud server equipped with four NVIDIA A100 Tensor Core GPUs and 112 CPU cores, which serve as the primary inference devices. The detailed specifications of these devices are provided in Table. \ref{tab:devices}. 
For the software environment, the cloud inference engine operates on the vLLM framework \cite{kwon2023efficient}, while the edge inference engine is implemented by PyTorch \cite{paszke2019pytorch} and Transformers \cite{wolf2020transformers}.

\begin{table}[htbp]
\centering
\captionsetup[table]{singlelinecheck=off}
\begin{threeparttable}
\begin{tabular}{ccccc}
 \toprule
  & \multicolumn{1}{c}{Device} & \multicolumn{1}{c}{Memory} & \multicolumn{1}{c}{Memory} & \multicolumn{1}{c}{AI} \\
  &  & \multicolumn{1}{c}{} & \multicolumn{1}{c}{bandwidth} & \multicolumn{1}{c}{performance} \\
  \midrule
    Edge & Jetson AGX Orin & 64GB & 204.8GB/s &  137.5 TFLOPS\\
    Cloud & NVIDIA A100 & 80GB & 1935GB/s &  624 TFLOPS\\
 \bottomrule
\end{tabular}
\end{threeparttable}
\caption{Physical devices.} 
\label{tab:devices}
\end{table}

\begin{table*}[htbp]
\centering
\begin{tabular}{
  >{\centering\arraybackslash}p{1.3cm}
  >{\centering\arraybackslash}p{1.cm}
  >{\centering\arraybackslash}p{0.6cm}
  >{\centering\arraybackslash}p{1.cm}
  >{\centering\arraybackslash}p{0.6cm}
  >{\centering\arraybackslash}p{1.cm}
  >{\centering\arraybackslash}p{0.6cm}
  >{\centering\arraybackslash}p{1.cm}
  >{\centering\arraybackslash}p{0.6cm}
  >{\centering\arraybackslash}p{1.cm}
  >{\centering\arraybackslash}p{0.6cm}
  >{\centering\arraybackslash}p{1.cm}
  >{\centering\arraybackslash}p{0.6cm}
}
\toprule
\multicolumn{1}{c}{\textbf{Methods}} & \multicolumn{2}{c}{Qwen2.5-72B-Instruct} & \multicolumn{2}{c}{Llama3-70B-Instruct} & \multicolumn{2}{c}{Qwen2.5-32B-Instruct} & \multicolumn{2}{c}{Llama3-8B-Instruct} & \multicolumn{2}{c}{Qwen2.5-7B-Instruct} & \multicolumn{2}{c}{Qwen2.5-1.5B-Instruct} \\
\cmidrule(r){2-3} \cmidrule(r){4-5} \cmidrule(r){6-7} \cmidrule(r){8-9} \cmidrule(r){10-11} \cmidrule(r){12-13}
& Throughput & Latency & Throughput & Latency & Throughput & Latency & Throughput & Latency & Throughput & Latency & Throughput & Latency \\
\midrule
Cloud-only & 14.891 & 138.62 & 16.33 & 121.54 & 32.13 & 72.32 & 75.51 & 28.57 & 88.33 & 30.88 & 148.12 & 23.71 \\
Edge-only & OOM & OOM & OOM & OOM & OOM & OOM & 6.03 & 804.21 & 6.68 & 801.23 & 21.20 & 210.38 \\
Routing & 14.86 & 145.04 & 13.79 & 143.94 & 30.04 & 88.57 & 69.55 & 74.75 & 69.55 & 68.66 & 133.31 & 41.28 \\
\textbf{\ouralg} & \textbf{21.24} & \textbf{97.34} & \textbf{25.98} & \textbf{75.15} & \textbf{34.81} & \textbf{61.22} & 70.48 & 30.21 & 84.98 & 31.78 & 140.86 & 26.19 \\
\bottomrule
\end{tabular}
\caption{Inference efficiency comparison. In the table, the models refer to those running in the cloud. For routing and \ouralg, the SLM at edge is any model with fewer parameters than the cloud model. Throughput is measured in the number of queries processed per minute (\#queries/min), and latency is measured in seconds (s). ``OOM" stands for ``Out of Memory".}
\label{table:inference}
\end{table*}

\begin{table*}[htbp]
\centering
\begin{tabular}{llccccccccccc}
\toprule
\multicolumn{1}{c}{\textbf{Methods}} & \multicolumn{1}{c}{\textbf{Metrics}} & \multicolumn{1}{c}{\textbf{Overall}} & \multicolumn{9}{c}{\textbf{Question categories}} \\
\cmidrule(r){4-13} 
 & & & generic & knowledge & roleplay & fermi & coding & math & writing & reasoning & stem & humanities\\
\midrule
\multirow{6}{*}{Cloud-only} & Overall score & \textcolor{blue}{\textbf{8.03}} & 7.90 & 8.20 & 8.12 & 8.20 & 8.00 & 8.50 & 7.81 & 8.33 & 8.20 &  7.83 \\
& Diversity rank& 1.82 & 1.90 & 1.80 & 1.59 & 1.50 & 1.50 & 1.36 & 1.56 & 1.67 & 2.50 & 2.78 \\
& Relevance rank& 1.67 & 1.40 & 1.00 & 1.06 & 1.10 & 1.00 & 1.00 & 1.17 & 1.00 & 4.00 & 4.00\\
& Immersion rank& 1.60 & 1.60 & 1.00 & 1.35 & 1.10 & 1.25 & 1.09 & 1.17 & 1.00 & 2.90 & 3.56\\
& Coherence rank& 1.55 & 1.30 & 1.00 & 1.18 & 1.00 & 1.25 & 1.00 & 1.17 & 1.17 & 3.60 & 2.89\\
& Integrity rank& 1.00 & 1.00 & 1.00 & 1.00 & 1.00 & 1.00 & 1.00 & 1.00 & 1.00 & 1.00 & 1.00\\
\midrule
\multirow{6}{*}{Edge-only} & Overall score & \textcolor{blue}{\textbf{7.38}} & 8.00 & 7.60 & 7.00 & 7.60 & 7.27 & 7.70 & 7.75 & 6.83 & 7.60 & 8.00\\
& Diversity rank& 1.81 & 1.70 & 1.60 & 1.88 & 1.40 & 1.81 & 2.18 & 2.28 & 2.67 & 1.40 & 1.22 \\
& Relevance rank& 2.01 & 1.20 & 1.00 & 2.24 & 1.60 & 2.63 & 3.36 & 2.50 & 3.50 & 1.10 & 1.00\\
& Immersion rank& 1.99 & 1.20 & 1.20 & 2.41 & 1.40 & 2.44 & 2.82 & 2.33 & 3.83 & 1.20 & 1.11 \\
& Coherence rank& 1.90 & 1.60 & 1.00 & 1.88 & 1.40 & 2.25 & 2.64 & 2.22 & 3.83 & 1.00 &  1.22\\
& Integrity rank& 1.44 & 1.00 & 1.00 & 1.38 & 1.00 & 1.56 & 2.36 & 1.83 & 2.33 & 1.00 & 1.00\\
\midrule
\multirow{6}{*}{Routing} & Overall score & \textcolor{blue}{\textbf{7.52}} & 7.70 & 7.70 & 7.82 & 7.90 & 7.67 & 8.30 & 8.00 & 6.67 & 7.80 & 7.33 \\
& Diversity rank& 1.81 & 2.00 & 2.10 & 1.76 & 1.50 & 1.94 & 2.09 & 1.67 & 2.33 & 1.40 & 1.56\\
& Relevance rank& 1.71 & 1.20 & 1.30 & 1.47 & 1.70 & 2.00 & 2.36 & 1.67 & 2.33 & 1.20 & 1.44 \\
& Immersion rank& 1.90 & 1.60 & 1.90 & 1.47 & 2.20 & 1.81 & 3.00 & 1.56 & 2.83 & 1.50 & 1.22\\
& Coherence rank& 1.75 & 2.40 & 1.00 & 1.24 & 1.90 & 2.06 & 2.36 & 1.28 & 2.83 & 1.20 & 1.44 \\
& Integrity rank& 1.32 & 1.00 & 1.00 & 1.00 & 1.30 & 1.31 & 1.27 & 1.00 & 2.00 & 1.60 & 1.33\\
\midrule
\multirow{6}{*}{\ouralg} & Overall score & \textcolor{blue}{\textbf{8.37}} & 8.30 & 8.60 & 8.65 & 8.60 & 7.00 & 8.10 & 8.12 & \textcolor{blue}{\textbf{9.00}} & \textcolor{blue}{\textbf{8.80}} & 8.50  \\
& Diversity rank& 1.70 & 1.40 & 1.70 & 2.00 & 2.10 & 1.44 & 1.18 & 1.39 & 2.00 & 2.20 & 2.44 \\
& Relevance rank& 1.10 & 1.00 & 1.00 & 1.12 & 1.10 & 1.00 & 1.00 & 1.11 & 1.67 & 1.50 & 1.56  \\
& Immersion rank& 1.21 & 1.20 & 1.00 & 1.59 & 1.10 & 1.13 & 1.09 & 1.11 & 1.67 & 1.20 & 1.00\\
& Coherence rank& 1.13 & 1.20 & 1.00 & 1.18 & 1.30 & 1.13 & 1.00 & 1.11 & 1.67 & 1.40 & 1.00\\
& Integrity rank& \textcolor{blue}{\textbf{1.00}} & 1.00 & 1.00 & 1.00 & 1.00 & 1.00 & 1.00 & 1.00 & 1.00 & 1.00 & 1.00 \\
\bottomrule
\end{tabular}
\caption{Response quality comparison. The overall scores are provided by FastChat \cite{zheng2024judging}, with values of (1-10), where higher scores indicate higher quality. In addition to a general metric, LLMZoo provides five detailed metrics \cite{LLMzoo}. For the same question, LLMZoo ranks the responses provided by the four methods across these metrics, with values of (1-4), where a rank of 1 indicates the best performance. Both evaluation frameworks use GPT-3.5 Turbo as the LLM judge.}
\label{table:llm_scores}
\end{table*}

\noindent\textbf{Metrics and benchmarks.} The performance of \ouralg~is evaluated using a series of Llama and Qwen models, as detailed in Table \ref{tab:ModelPerformance}. The key metric for inference efficiency is throughput, defined as the number of requests processed by the system per minute. Additionally, we measure average end-to-end latency, which encompasses inference latencies at both cloud and edge levels, waiting time, and data transfer latency. To evaluate response quality, we adopt two LLM-based evaluation frameworks: FastChat \cite{zheng2024judging} and LLMZoo \cite{LLMzoo}. These frameworks involve presenting a series of open-ended questions to an LLM judge (e.g., GPT-3.5 turbo) to score and rank responses based on quality. We specifically utilize two datasets, MT-bench and Vicuna-bench \cite{zheng2024judging}, designed to assess the instruction-following abilities of chatbots and differentiate them based on their core capabilities (\eg, reasoning and mathematics).

\noindent\textbf{Baselines.} We compare the performance in terms of inference efficiency
and answer quality of \ouralg~with the baselines.
\begin{itemize}
    \item Cloud-only. All queries are serviced by language models deployed on the cloud with the vLLM framework \cite{kwon2023efficient}.
    \item Edge-only. In this case, all queries are handled by SLMs ($\le$ 8B) that are locally deployed on edge devices. Each edge device hosts a model, and queries are dispatching in a load-balanced manner.
    \item Routing \cite{ding2024hybrid}. It uses a router to dynamically allocate queries to either a small at edge or large model at cloud based on the predicted query difficulty.
\end{itemize}

\subsection{Overall Performance}\label{sec:overall_performance}

\noindent\textbf{Throughput and latency.} To assess throughput, we configure the cloud server's maximum batch size for the Qwen2.5-72B-Instruct model at 20, with other devices and models proportionally adjusted. The requests per minute (RPM) is 1.5 times the maximum batch size. The throughput and end-to-end latency of LLM inference are shown in Table. \ref{table:inference}. We have the following observations. First, \ouralg~is potential and beneficial for large language model deployment. \ouralg~achieves significantly higher inference throughput and lower inference latency than baseline methods by leveraging edge computing resources. For the Llama3-70B model, \ouralg~delivers an average throughput of 25.98 requests per minute, approximately 1.59 times that of Cloud-only and 2.32 times that of the Routing method, while also reducing latency by about 38\% to 58\%. Similar improvements are noted for the Qwen2.5-72B model. Secondly, for the Qwen2.5-32B model, the performance of \ouralg~is on par with Cloud-only. This is attributed to the model's poor ability to predict response lengths, often underestimating them, leading \ouralg~to favor not engaging the progressive inference mode. Thirdly, for smaller models such as the Llama3-8B, the efficiency of \ouralg~is slightly inferior to Cloud-only but superior to both Routing and Edge-only. This is because the size difference between small edge models and cloud models is insufficient, making edge inference a bottleneck for the system. Therefore, we recommend that the size of the LLM in \ouralg~should be more than ten times that of the SLM.

\noindent\textbf{Response quality.} As shown in Table \ref{table:llm_scores}, \ouralg~achieves higher overall scores than the baselines, indicating that \ouralg, due to its integration of multiple models' outputs, improves response quality compared to using an LLM alone, particularly for knowledge, roleplay, and reasoning questions. However, \ouralg~performs less well on math and coding tasks than the Cloud-only method, due to the difficulty in fully capturing essential meanings in the preliminary sketches generated by the LLM. Additionally, \ouralg~outperforms the baseline on coherence, diversity, immersion, integrity, and relevance metrics, achieving the best average ranking, especially in terms of integrity across all tasks. This performance suggests that the two-stage approach of progressive inference, while reliant on the quality of LLM-generated sketches, effectively maintains completeness and promotes diversity in the outputs.

\subsection{Performance Breakdown}\label{sec:performance_breakdown}

We evaluate the benefits brought by \ouralg's key designs separately. The experiments are performed with LLama3-70B-Instruct in the cloud, and the three SLMs in the edge.

\noindent\textbf{Efficiency improvement with dynamic scheduler.}
\begin{figure*}[ht] 
  \centering
  \begin{subfigure}{0.32\textwidth}
    \includegraphics[width=\linewidth]{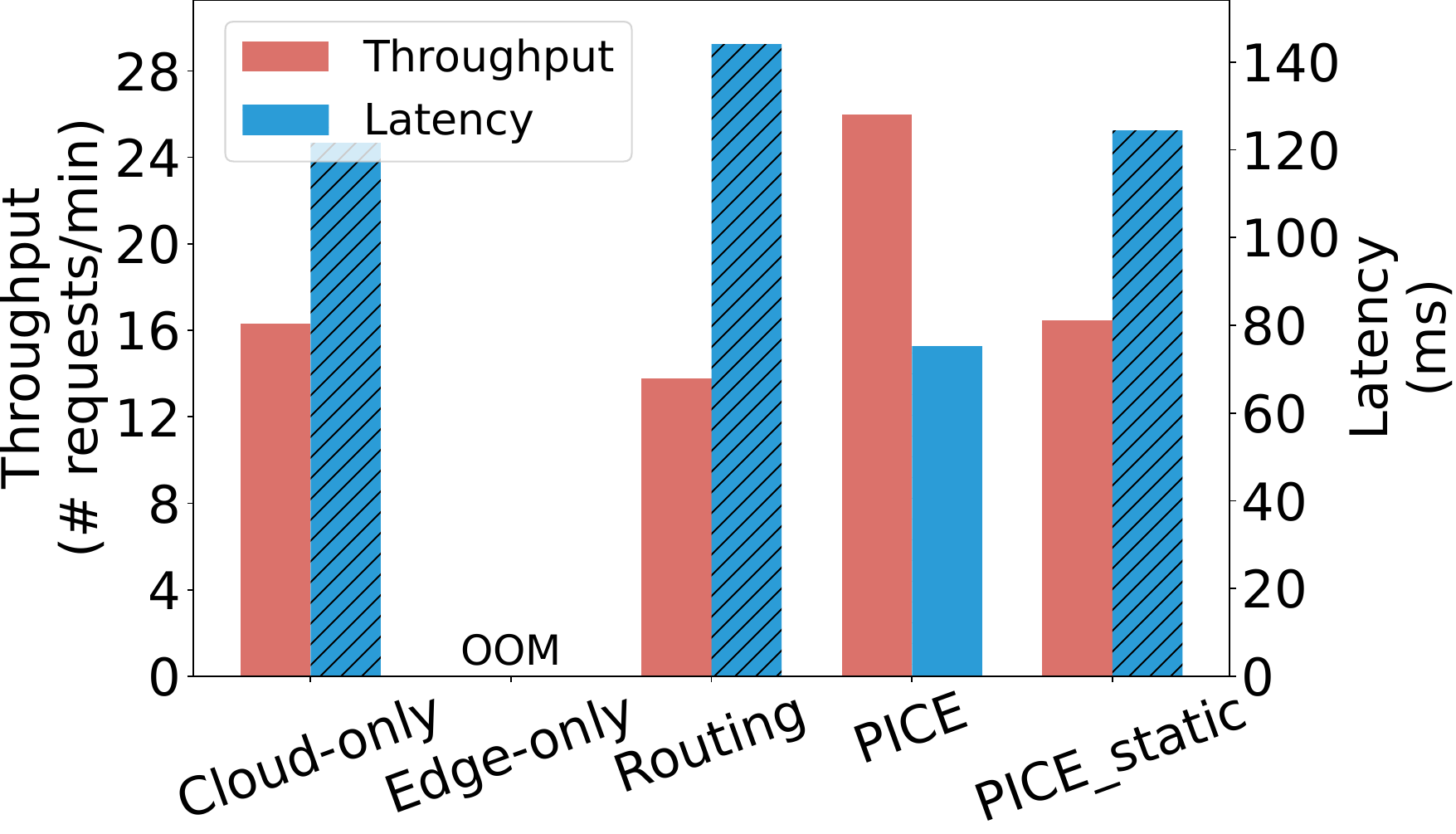}
    \caption{The throughput and latency.}
    \label{fig:lab31_1}
  \end{subfigure}
    \begin{subfigure}{0.31\textwidth}
    \includegraphics[width=\linewidth]{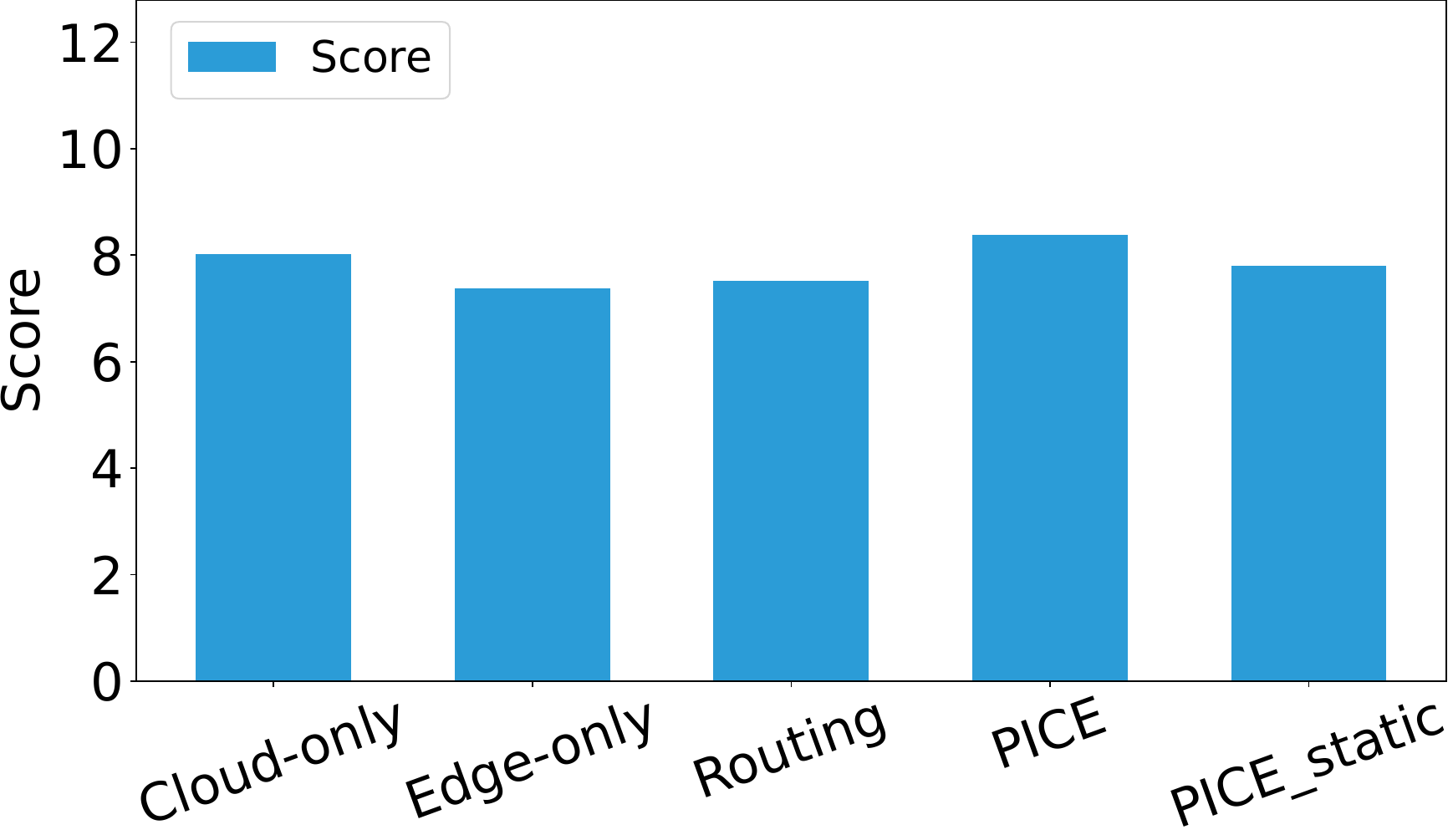}
    \caption{The response quality.}
    \label{fig:lab31_2}
   \end{subfigure}
    \begin{subfigure}{0.30\textwidth}
    \includegraphics[width=\linewidth]{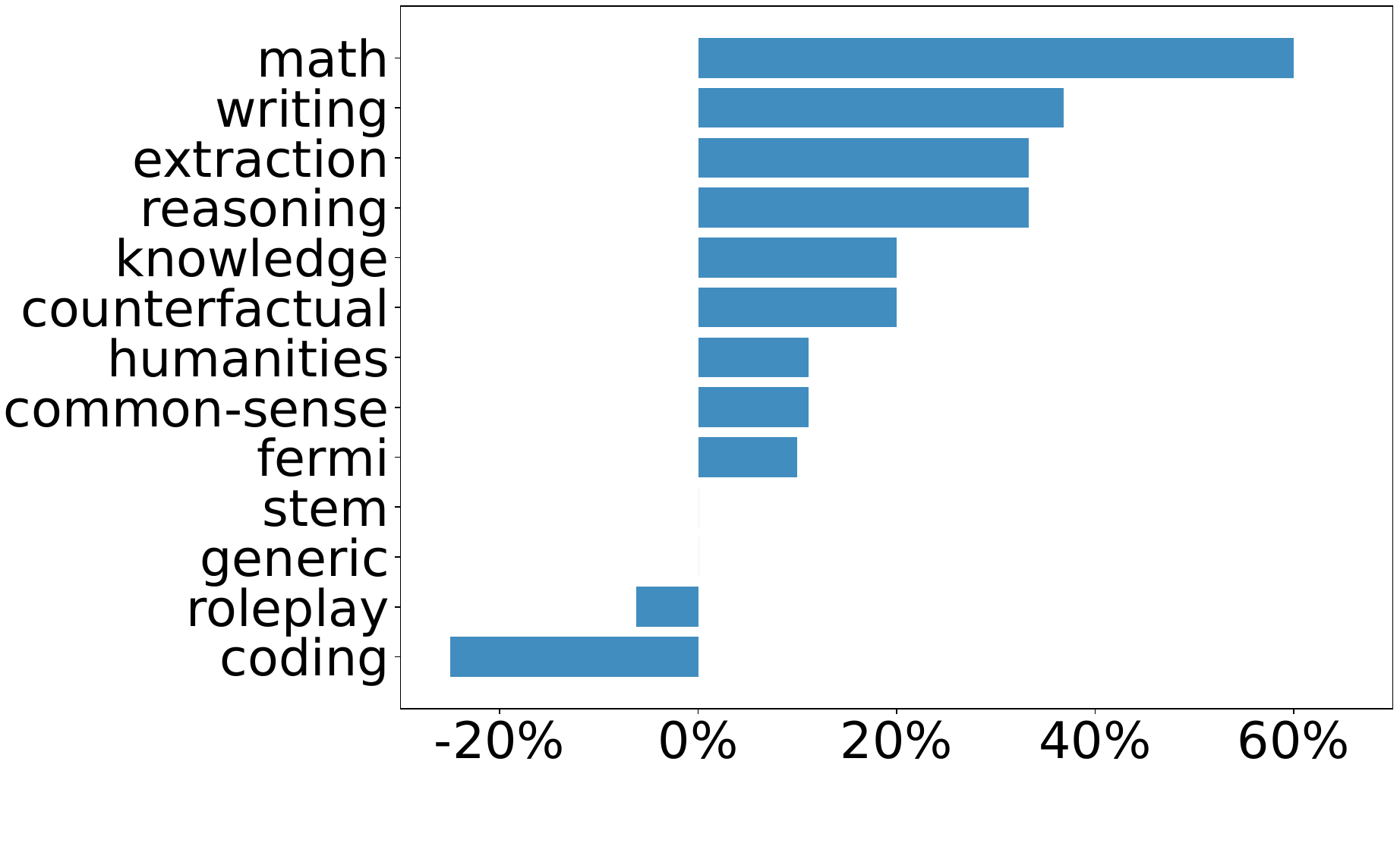}
    \caption{Net win rate.}
    \label{fig:lab31_3}
  \end{subfigure}
  \caption{The efficiency and response quality improvement with dynamic scheduler. Net win rate in (c) is the difference between the fraction of questions that \ouralg~has better and worse answers than \ouralg~with static scheduler.}
  \label{fig:lab31_and_2}
\end{figure*}
The dynamic scheduler in \ouralg~balances inference efficiency and response quality by adapting to runtime conditions. In contrast, static scheduling adheres to predefined rules and parameters, allocating requests between cloud and edge devices based solely on predicted answer lengths. As depicted in Fig. \ref{fig:lab31_and_2}(a), while \ouralg~with static scheduling modestly improves throughput compared to the Cloud-only method, dynamic scheduling significantly enhances throughput by an additional 50\%. Additionally, dynamic scheduling significantly reduces latency compared to static scheduling and other methods like Cloud-only or Routing, demonstrating its effectiveness in balancing throughput with response time.

As shown in Fig.~\ref{fig:lab31_and_2}(b), response quality is improved by 6.0\%, even surpassing that of standalone LLMs, as \ouralg~produces more detailed and comprehensive answers. As detailed in Fig. \ref{fig:lab31_and_2}(c), the dynamic scheduler markedly enhances answer quality in 69\% of question categories, with significant improvements in mathematics and writing tasks, where it achieves net win rates of 60\% and 36.8\%, respectively.

\begin{figure}[htbp]
  \centering
  \begin{subfigure}{0.49\columnwidth}
        \includegraphics[width=0.99\columnwidth]{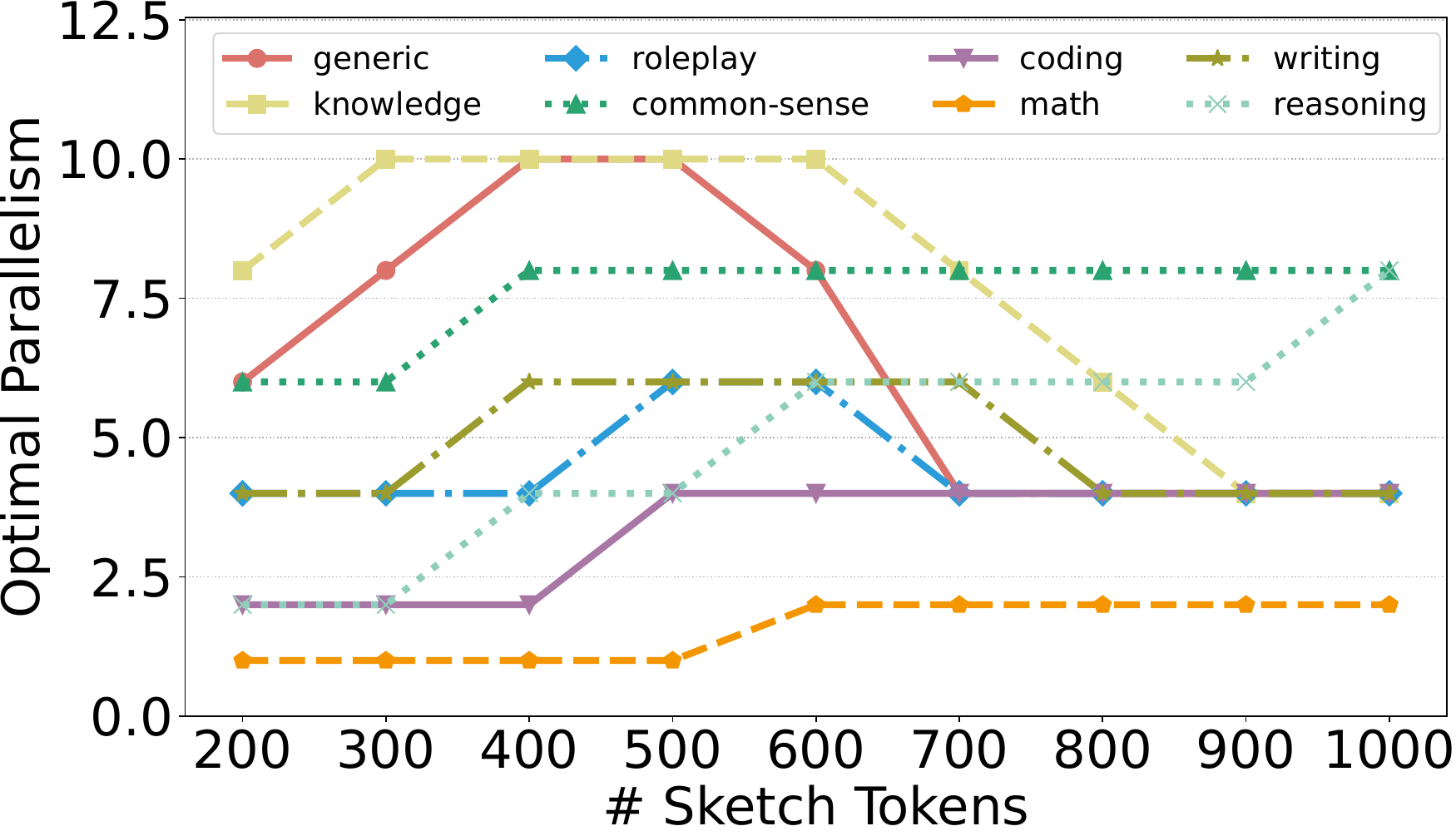} 
        \caption{The optimal parallelism.}
        \label{fig:lab32_1}
  \end{subfigure}
  \begin{subfigure}{0.49\columnwidth}
        \includegraphics[width=0.99\columnwidth]{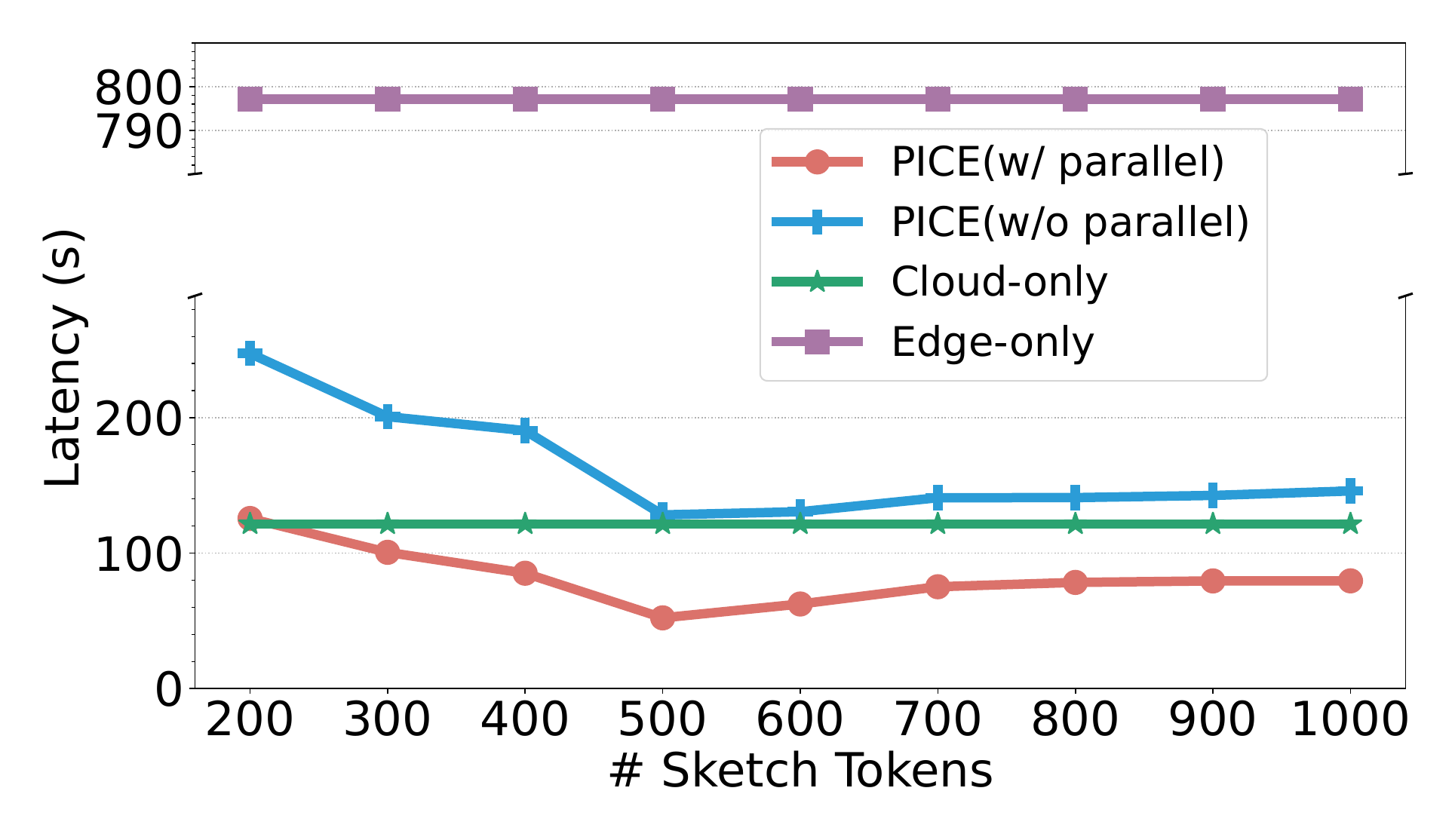} 
        \caption{The latency of algorithms.}
        \label{fig:lab32_2}
  \end{subfigure}
  \caption{The impact of parallel generating.}
  \label{fig:lab_32}
\end{figure}

\noindent\textbf{Latency reduction with execution optimizer.}
The execution optimizer capitalizes on semantic-level parallelism to accelerate edge inference. In Fig.~\ref{fig:lab_32}(a), we analyze the optimal parallelism of different tasks based on the sketch length. For tasks such as generic and roleplay,  parallelism increases with sketch length, peaks around 500 tokens, and then declines, indicating benefits from longer sketches until limited by edge device memory. Conversely, tasks like common-sense and math maintain consistently low parallelism, as their inherently shorter responses do not require high parallelism.

As depicted in Fig. \ref{fig:lab_32}(b), the parallel mechanism decreases the inference latency of the \ouralg~system at least 62s. When the sketch token count is below 500, increasing the length of sketches enhances the parallelism in edge inference, significantly reducing latency. However, once the count exceeds 500 tokens, the latency reduction with \ouralg~becomes less marked. This occurs because the required KV cache for the prompts enlarges, and the limited memory of edge devices cannot support further increases in parallelism. These results emphasize the need to adjust parallelism based on task type and sketches to optimize inference efficiency.

\begin{figure}[htbp]
    \centering
    \includegraphics[width=0.8\columnwidth]{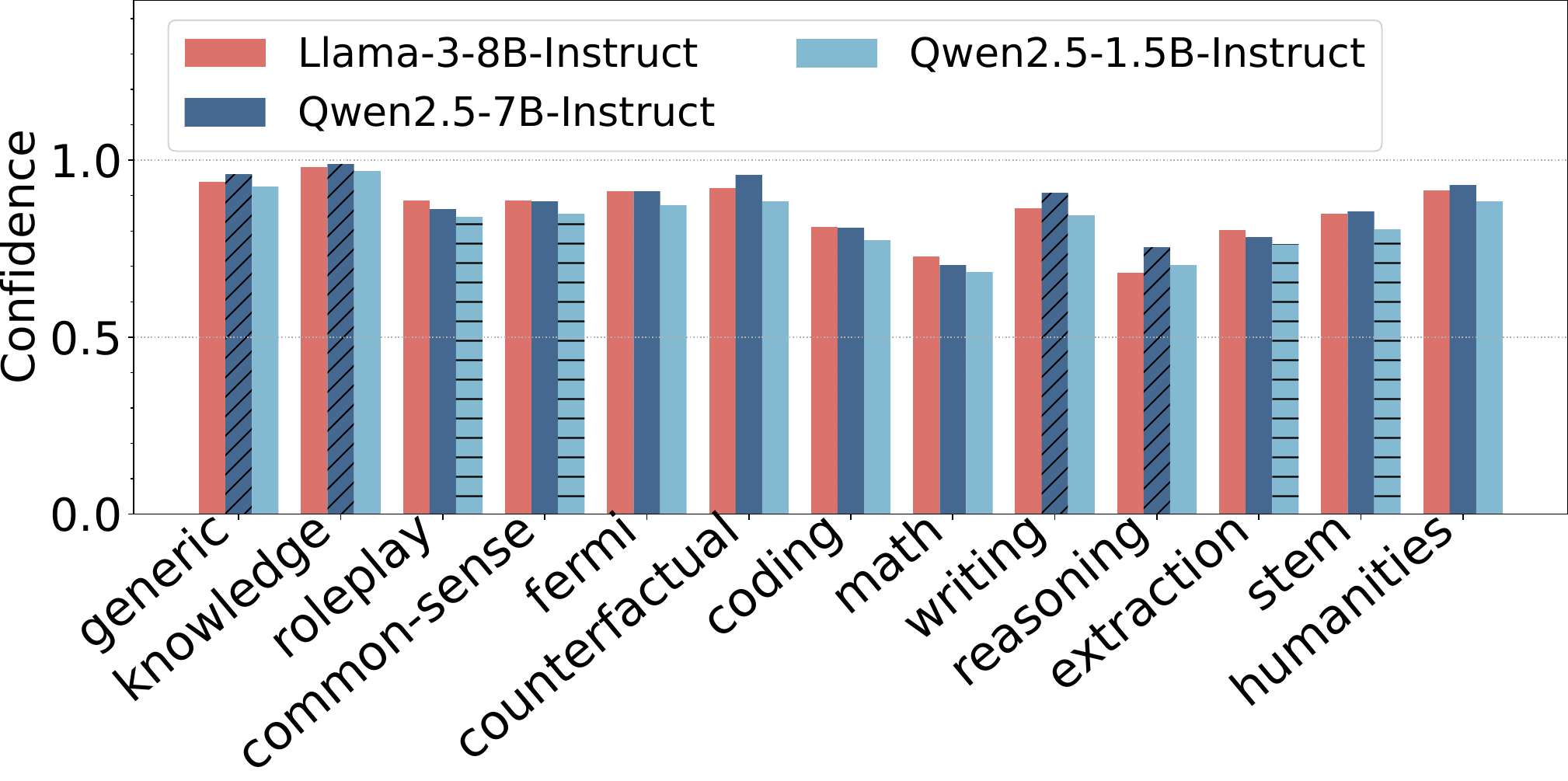}
    \caption{The confidence of SLMs.}
    \label{fig:lab33_1}
\end{figure}

\begin{figure}[htbp]
    \centering
    \includegraphics[width=0.8\columnwidth]{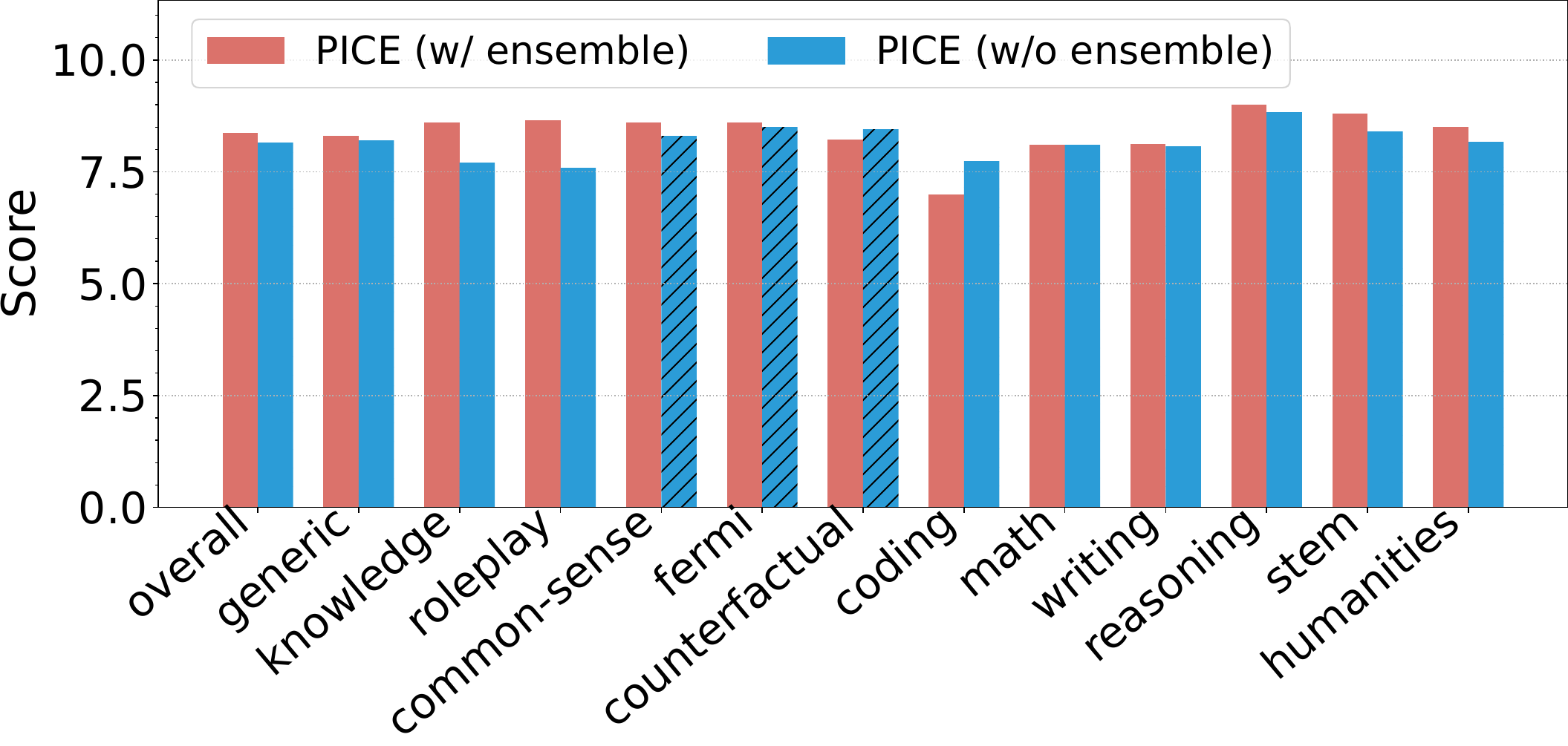}
    \caption{Impact of ensemble learning on response quality.}
    \label{fig:lab33_2}
\end{figure}

\noindent\textbf{Quality improvement with ensemble learning.}
As shown in Fig. \ref{fig:lab33_1}, the confidence rankings of SLMs differ significantly across various question categories, underscoring each model's specialized capabilities. Leveraging this diversity, \ouralg~adopts an ensemble learning approach to select the best answer. As evidenced by Fig. \ref{fig:lab33_2}, this method enhances response quality in nearly all categories, except for coding and counterfactual scenarios, with the most substantial improvements noted in roleplay (13.95\%) and knowledge (11.69\%).

Furthermore, the overall response quality is improved by 2.8\%, demonstrating the effectiveness of ensemble learning in combining the strengths of individual models. The notable gains in categories like roleplay and knowledge highlight the ensemble's ability to better capture nuanced and detailed information, ensuring higher-quality outputs in tasks that benefit from diverse model capabilities.

\begin{figure}[htbp]
    \centering
    \includegraphics[width=0.8\columnwidth]{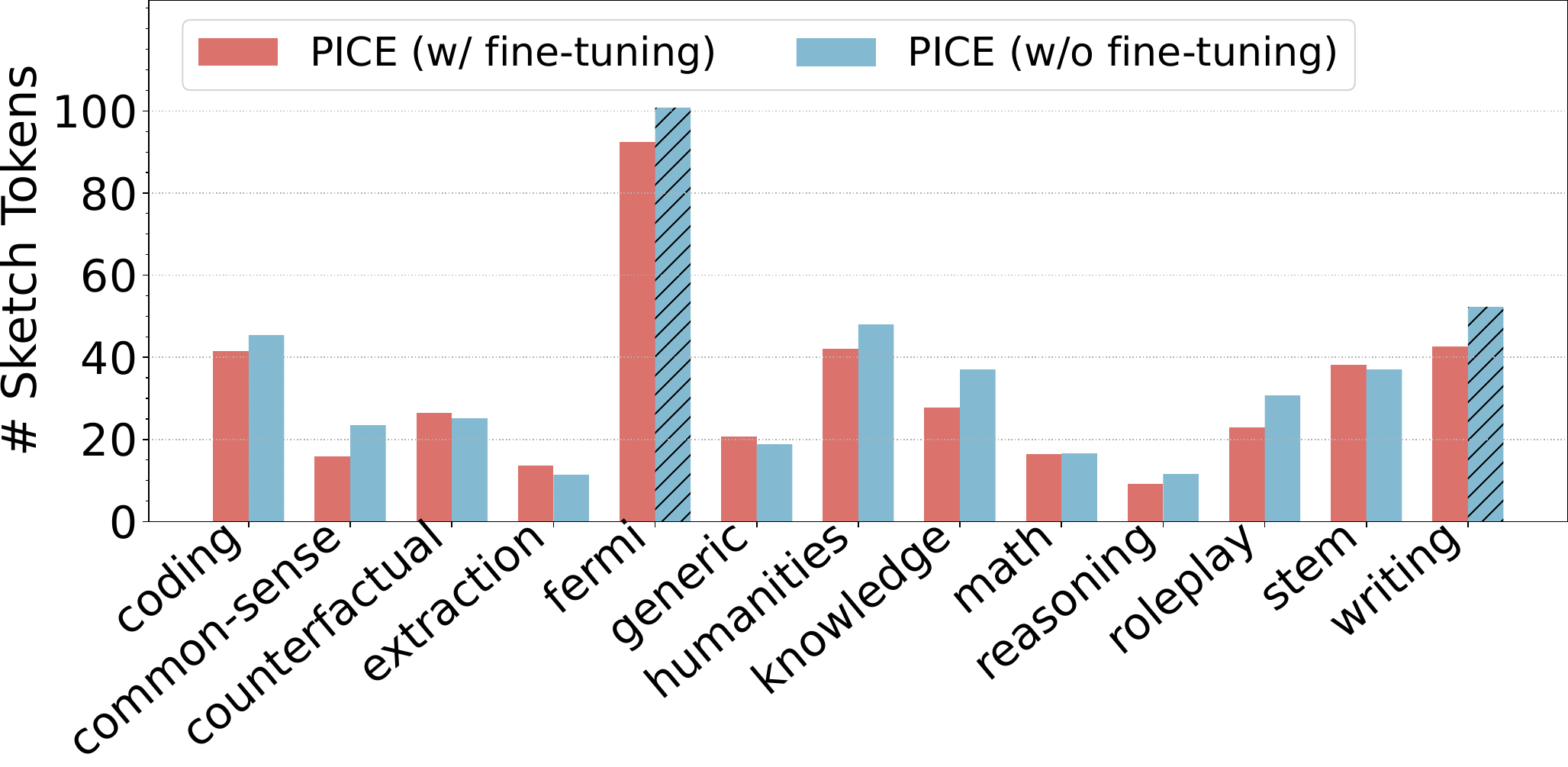}
    \caption{The length of sketches across categories.}
    \label{fig:lab30_1}
\end{figure}

\noindent\textbf{Performance improvement with model Fine-tuning.} The experiments demonstrate the impact of the fine-tuning component on the model’s ability to generate concise and semantically complete sketches. As shown in Fig.~\ref{fig:lab30_1}, compared to the base model, the fine-tuned model achieves a reduction in sketch length in several categories. For instance, the writing category decreases significantly from 52.3 to 42.6, and knowledge from 36.9 to 27.7. In contrast, slight increases in categories like counterfactual (from 25 to 26.4) and generic (from 18.7 to 20.7) suggest that the fine-tuning process preserves or even enhances performance in these specific areas, likely by generating longer sketches to express semantics.

\begin{figure}[htbp]
    \centering
    \includegraphics[width=0.8\columnwidth]{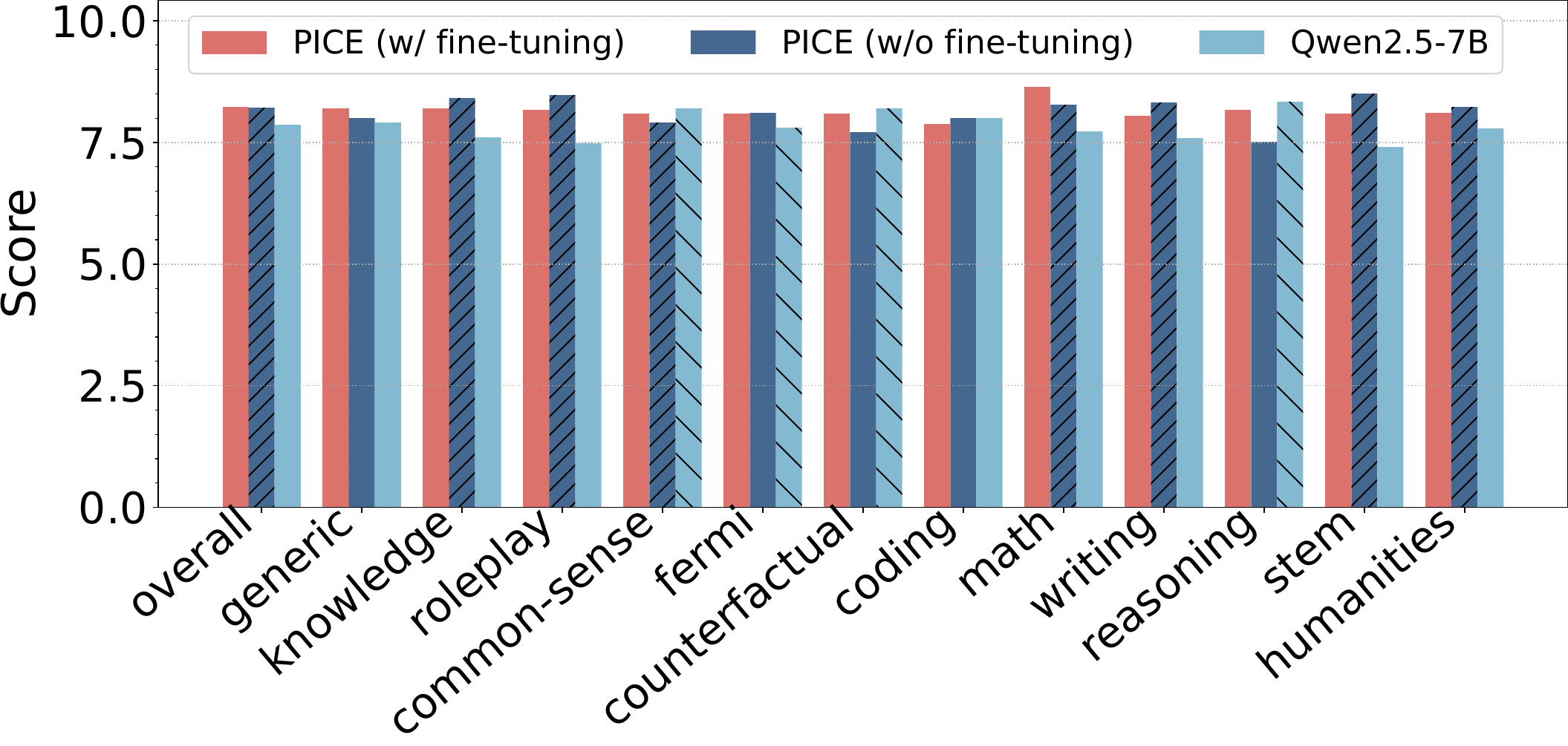}
    \caption{Impact of the fine-tuning component on response quality across categories.}
    \label{fig:lab30_2}
\end{figure}

The results shown in Fig.~\ref{fig:lab30_2} illustrate the model's response quality before and after applying reinforcement learning from AI feedback. The results demonstrate the algorithm's effectiveness, with the fine-tuned model achieving higher scores in categories like generic, math, and reasoning, indicating improved response quality. However, in knowledge and writing, the fine-tuned model performs slightly worse, likely due to its focus on generating concise sketches, which can occasionally sacrifice critical semantic details. This highlights a trade-off between conciseness and completeness, especially for tasks needing detailed explanations.

\subsection{Sensitivity Analysis}\label{sec:sensitivity_analysis}
We conduct experiments under varying conditions, including different requests per minute (RPM), job queue lengths, and network environments, to comprehensively evaluate the robustness and adaptability of \ouralg. 

\begin{figure}[htbp]
  \centering
  \begin{subfigure}{0.49\columnwidth}
        \includegraphics[width=0.99\columnwidth]{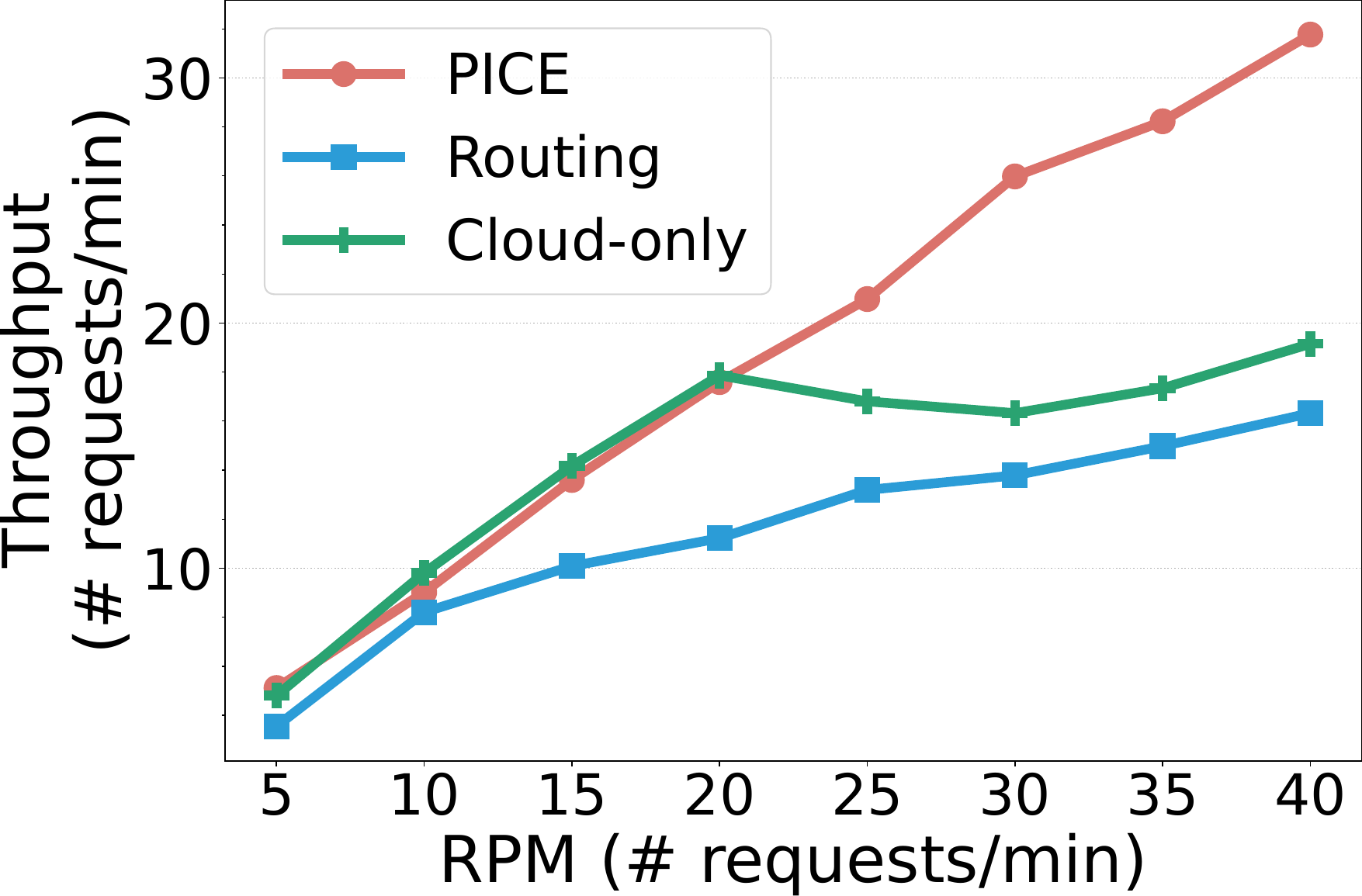} 
        \caption{Throughput.}
        \label{fig:lab41_1}
  \end{subfigure}
  \begin{subfigure}{0.49\columnwidth}
        \includegraphics[width=0.99\columnwidth]{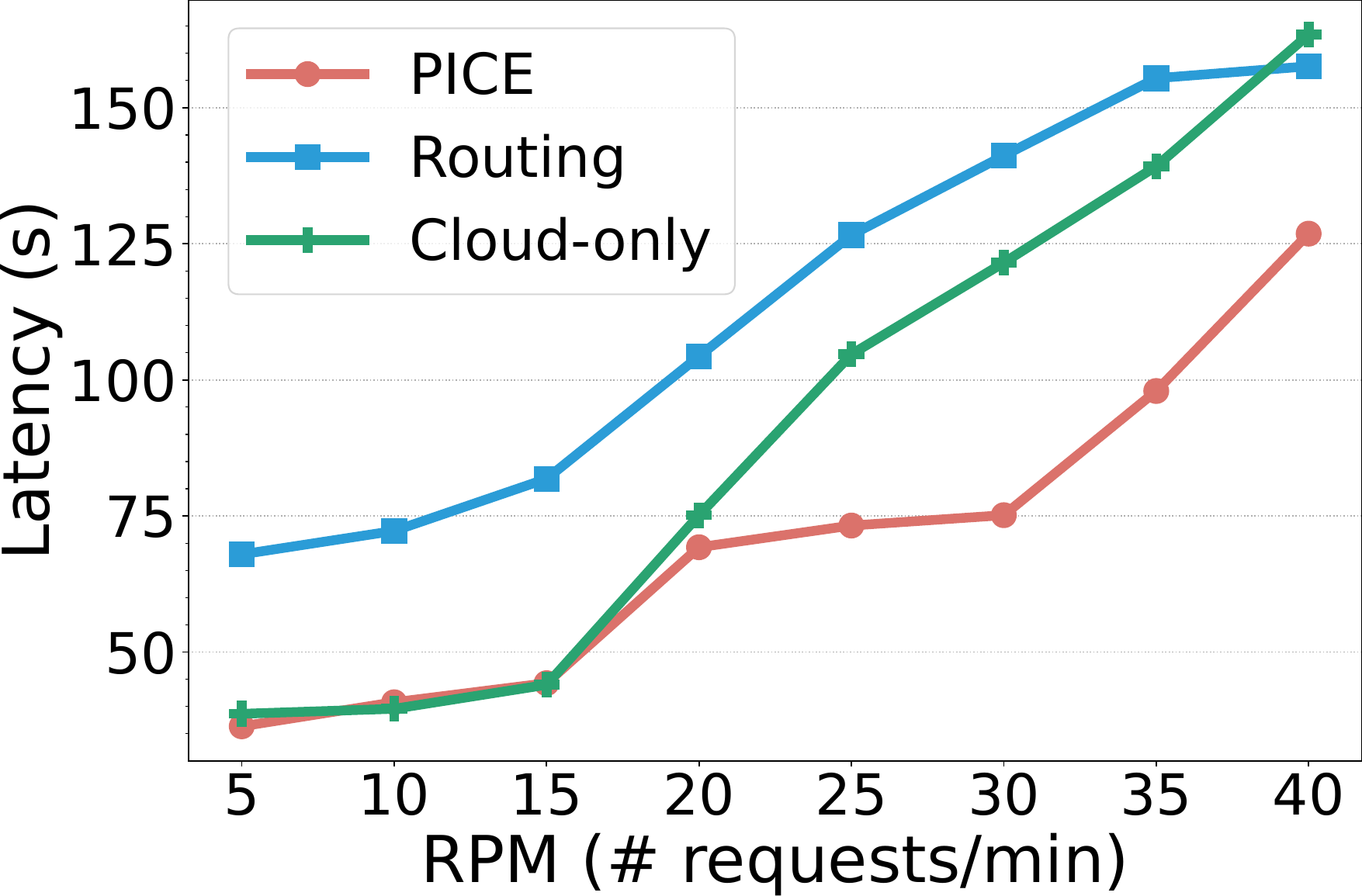}
        \caption{Latency.}
        \label{fig:lab41_2}
  \end{subfigure}
  \caption{The impact of RPM (requests per minute).}
  \label{fig:lab_41}
\end{figure}

\noindent\textbf{Impact of RPM.} Fig.\ref{fig:lab_41} illustrates the effect of varying RPM on system performance. When the RPM is below 20, which corresponds to the maximum batch size supported by the LLM in the cloud, the performance of \ouralg~is comparable to the Cloud-only approach. However, as the RPM exceeds 20, \ouralg~continues to enhance throughput by offloading tasks to edge devices, whereas the throughput of Cloud-only remains static and its end-to-end latency increases sharply due to longer wait times. While the Routing method attempts to dynamically allocate requests to edge devices, its efficiency is limited by the constrained resources available at the edge.

\begin{figure}[htbp]
    \centering
    \includegraphics[width=0.8\columnwidth]{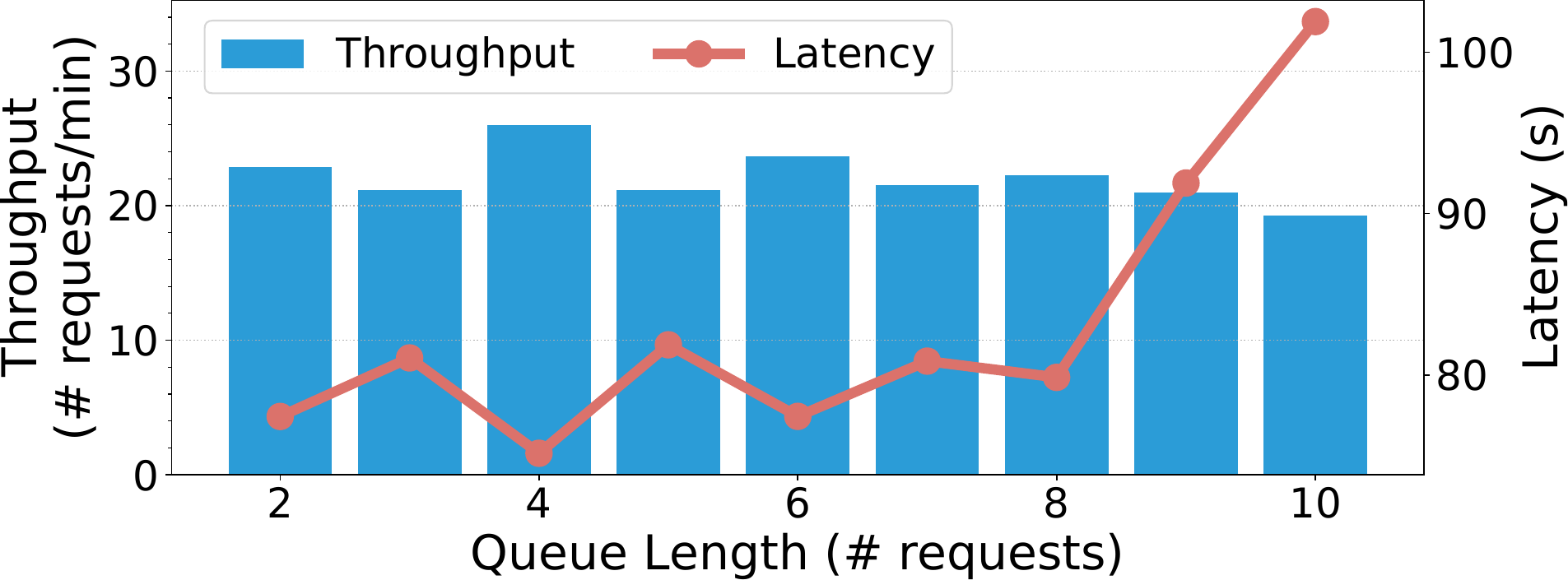}
    \caption{The impact of job queue length.}
    \label{fig:lab42}
\end{figure}

\noindent\textbf{Impact of the job queue.} The configuration of the job queue significantly affects \ouralg's scheduling strategy. As shown in Fig.~\ref{fig:lab42},optimal throughput is achieved when the queue length is set to four, allowing each edge device to handle one pending request while the cloud-based LLM manages the remaining workload. When the queue length exceeds eight, waiting times increase significantly, leading to a substantial rise in overall latency. This highlights the importance of aligning job queue parameters with the number of edge devices and the latency associated with edge inference. Proper tuning of these parameters ensures efficient resource utilization and minimizes delays, enhancing the overall system performance.

\begin{figure}[htbp]
  \centering
  \begin{subfigure}{0.49\columnwidth}
        \includegraphics[width=0.99\columnwidth]{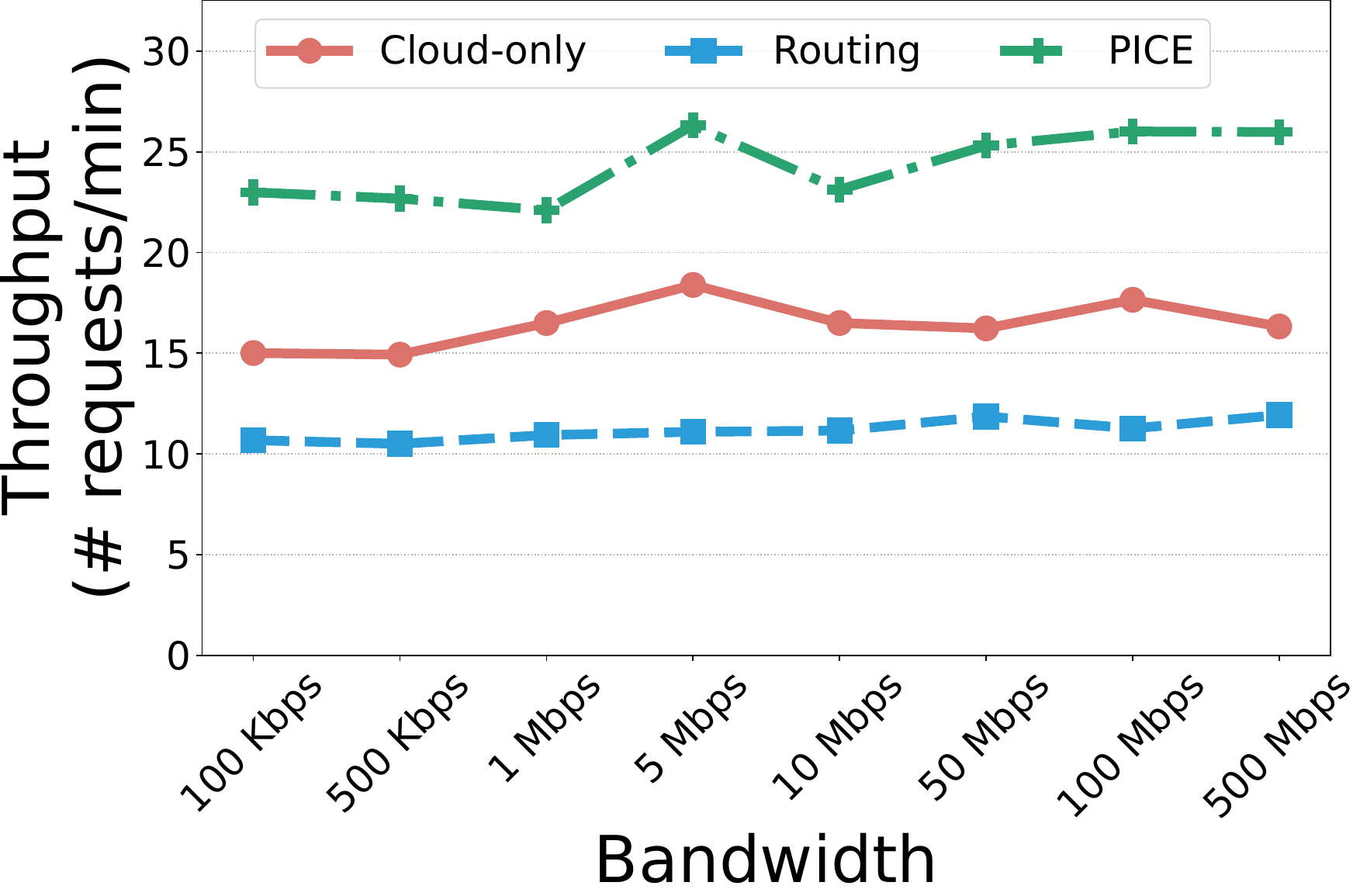} 
        \caption{Throughput.}
        \label{fig:lab43_1}
  \end{subfigure}
  \begin{subfigure}{0.49\columnwidth}
        \includegraphics[width=0.99\columnwidth]{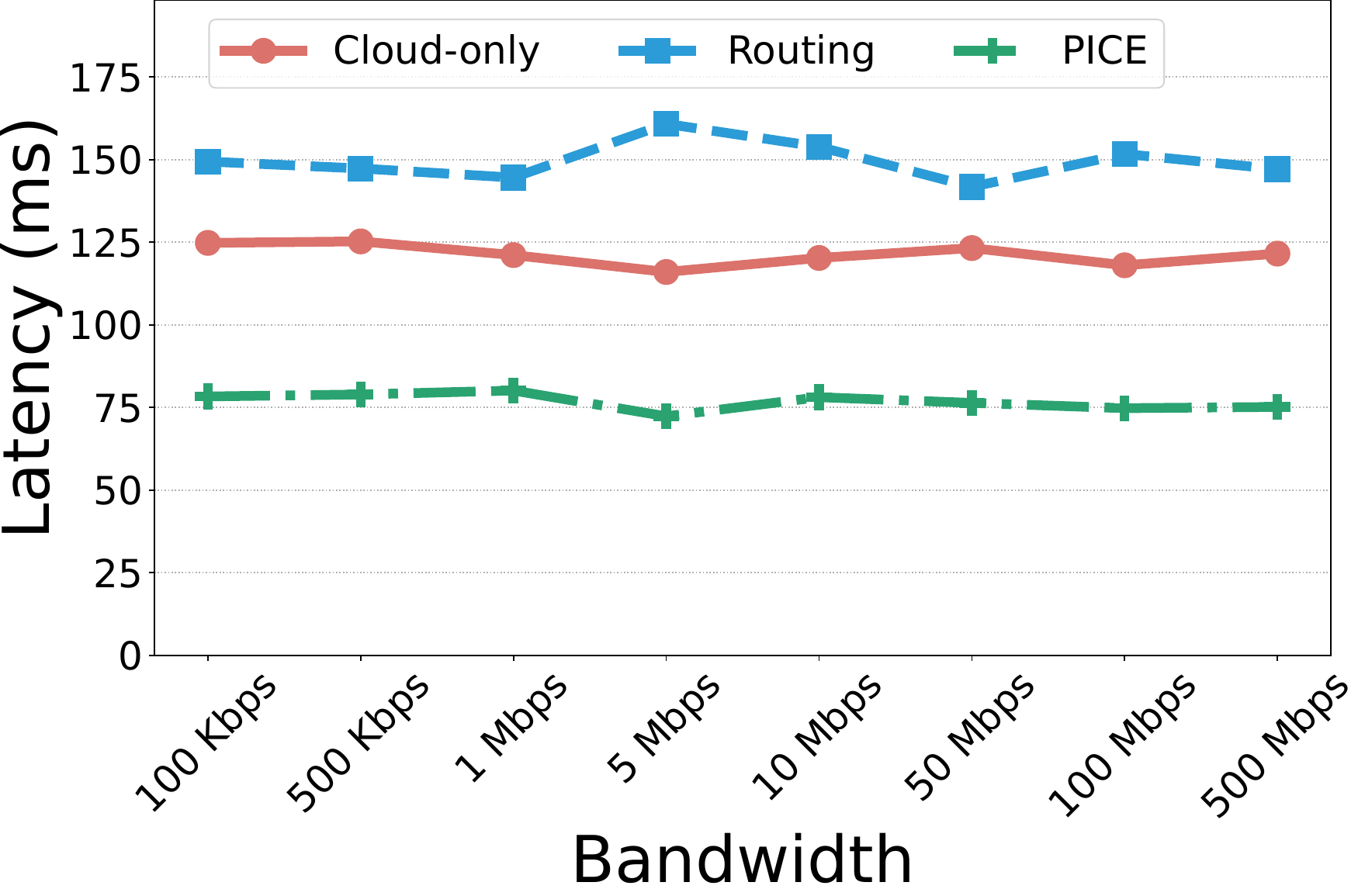}
        \caption{Latency.}
        \label{fig:lab43_2}
  \end{subfigure}
  \caption{The impact of bandwidth on inference efficiency.}
  \label{fig:lab_43}
\end{figure}

\noindent\textbf{Impact of bandwidth.} Fig.~\ref{fig:lab_43} illustrates the effects of varying bandwidth between cloud and edge devices on the latency and throughput of the LLM inference system. In Fig.~\ref{fig:lab_43}, throughput is analyzed under different bandwidth conditions. The PICE approach demonstrates consistently higher throughput compared to both Cloud-only and Routing methods across all bandwidth levels. This indicates the efficiency of PICE in leveraging edge resources to offload tasks and maintain performance. The Cloud-only method shows limited improvement with increasing bandwidth, highlighting its reliance on centralized resources. Meanwhile, the Routing approach underperforms due to suboptimal task allocation.

Fig.~\ref{fig:lab_43} shows that PICE achieves the lowest latency across all bandwidths due to dynamic scheduling and task offloading. Across all methods, including PICE, Cloud-only, and Routing, latency shows minimal variation as bandwidth increases. This is because the primary computational overhead lies in inference, while data transmission between the cloud and edge devices—limited to queries or sketches—takes only a few tens of milliseconds, even at lower bandwidths. These results highlight that network bandwidth has minimal impact on overall system performance, with inference time being the dominant factor.
\section{Related Work}
We review research on LLMs from the areas of response quality, inference efficiency, and the balance between the two.

\noindent\textbf{Response quality.} Ning \etal~propose SoT \cite{ning2024skeleton}, guiding LLMs to first generate the skeleton of the answer and then filling in the content through parallel API calls. However, the applicability of SoT is constrained by its parallel mechanism. When queries involve a progressive relationship between points, such as creating a travel guide detailing Paris attractions and food along the way, SoT may encounter limitations in providing valid answers. 
Additionally, ensemble models have been shown to significantly enhance response quality in multi-task learning and the processing of complex data. For instance, \cite{shen2023mixture} introduced an approach that integrates instruction tuning with sparse MoE techniques, effectively augmenting the expressive capacity of language models while concurrently improving both accuracy and computational efficiency.
Inspired by these works, we use LLM to generate semantically complete short sentences and then use SLM to enrich each sentence, alleviating the limitations of the single model thinking paradigm.

\noindent\textbf{Inference efficiency.} At the system level, the speculative decoding method \cite{miao2024specinfer, xu2023llmcad, leviathan2023fast} adopts an SLM to generate some consecutive tokens in sequence and applies the target LLM to verify them in parallel to break the sequential decoding of LLM. Kwon \etal~propose a novel memory management mechanism called PagedAttention and establish a high throughput system vLLM \cite{kwon2023efficient}. There are some other works that explore the distributed deployment of LLMs to enhance efficiency, such as EdgeShard\cite{zhang2024edgeshard} and PrivateLoRA\cite{wang2023privatelora}. Specifically, \cite{zhang2024edgeshard} proposes the EdgeShard framework, which partitions large language models (LLMs) into multiple segments and leverages dynamic programming algorithms to optimize their deployment across distributed devices, thereby enhancing the efficiency of LLM inference. At the model level, methodologies such as model quantization \cite{kim2023squeezellm, huang2024good, xu2023qa, liu2024kivi}, pruning \cite{guo2023compresso}, and distillation \cite{yang2020model} are employed to achieve model lightweighting. These approaches aim to reduce the model's parameter size and computational complexity while preserving its performance as much as possible, thereby facilitating deployment in resource-constrained environments. \cite{rajbhandari2022deepspeed} introduces DeepSpeed-MoE. By incorporating a MoE architecture design and utilizing model compression techniques, such as knowledge distillation and slicing, the system reduces the size of the MoE model by a factor of 3.7.

\noindent\textbf{The balance between the efficiency and quality.} Tabi integrates SLMs with the LLMs to reduce inference latency while maintaining high accuracy for classification tasks\cite{wang2023tabi}. For generative tasks, a hybrid inference strategy is proposed, which uses routers to dynamically assign queries to small models at the edge or large models in the cloud according to the predicted query difficulty \cite{ding2024hybrid, ong2024routellm, xu2024conveyor}. DejaVu uses the input of each layer of to predict context sparsity, and adopts asynchronous and hardware-aware methods to speed up inference while ensuring model quality\cite{liu2023deja}.

\section{Conclusion}

We propose \ouralg, a progressive inference system for LLM serving over cloud and edge, to effectively balance efficiency and response quality. We ensure the robust performance of \ouralg~through a dynamic scheduler that leverages edge computing to optimize throughput, an ensemble learning mechanism that aggregates outputs from multiple small models to enhance quality, an execution optimizer that employs semantic-level parallelism to accelerate edge inference and a model fine-tuning strategy that optimizes LLMs for generating concise and semantically complete sketches. Experiments demonstrate that \ouralg~achieves a $1.5- 2\times$ increase in throughput and a $43\%$ reduction in latency without compromising response quality, providing a viable solution for the widespread deployment of LLM applications. \ouralg~is an initial attempt at semantic-centric optimization for inference efficiency, and showcases the potential of eliciting high-quality answers by explicitly planning the answer structure.


\bibliographystyle{IEEEtran}
\bibliography{ref}



 




\vfill

\end{document}